\documentclass{aa}  

\usepackage{CJK}  
\usepackage{enumitem}
\usepackage{graphbox}
\usepackage{soul}

\usepackage{graphicx}
\usepackage{txfonts}
\usepackage{lipsum}
\usepackage{lscape}
\usepackage{placeins}

\usepackage[colorlinks=true,      
            citecolor=blue,       
            linkcolor=blue,      
            urlcolor=blue,       
            bookmarks=true,       
            breaklinks=true]{hyperref} 

\usepackage{multirow}

\begin{document}

\begin{CJK}{UTF8}{gbsn}  

   \title{Observational evidence of the Gaia-Enceladus-Sausage merger's influence on the opposite side of the Galactic Center}

\author{Yuxiao Chen (陈昱晓)\inst{1,2}
    \and Gaochao Liu (刘高潮)\inst{1,2}\fnmsep\thanks{Corresponding author: gcliu@ctgu.edu.cn}
        \and Yang Huang (黄样)\inst{3,4}\fnmsep\thanks{Corresponding author: huangyang@ucas.ac.cn}
        \and  Huawei Zhang (张华伟)\inst{5,6} 
        \and Timothy C. Beers \inst{7} 
        \and Dejian Liu (刘德剑)\inst{1,2}
        \and Sarah A. Bird\inst{1,2} 
        \and Peng Zhang (张鹏)\inst{1,2} 
        \and Yiping Zhang (张仪萍)\inst{1,2}
        }

\authorrunning{Chen et al.}

\institute{College of Science, China Three Gorges University, Yichang 443002, P. R. China
   \and Center for Astronomy and Space Sciences, China Three Gorges University, Yichang 443002, P. R. China
   \and School of Astronomy and Space Science, University of Chinese Academy of Sciences, Beijing 100049, P. R. China
   \and CAS Key Lab of Optical Astronomy, National Astronomical Observatories, University of Chinese Academy of Sciences, Beijing 100101, P. R. China
   \and Department of Astronomy, School of Physics, Peking University, Beijing 100871, P. R. China
   \and Kavli Institute for Astronomy and Astrophysics, Peking University, Beijing 100871, P. R. China
   \and Department of Physics and Astronomy and JINA Center for the Evolution of the Elements (JINA-CEE), University of Notre Dame, Notre Dame, IN 46556, USA
   }

\abstract
  {The Gaia-Enceladus-Sausage (GES) merger event, which occurred approximately 8 -- 11~Gyr ago, is believed to be a key event that shaped the current structure of the Galactic inner halo. Although the remnants of this event have been extensively studied and confirmed to dominate the chemical and dynamical properties of the halo on the Near-side of the Galactic Center (GC) (i.e., the side facing the Sun), systematic observational evidence for whether the Far-side of the GC (i.e., the side beyond the GC as seen from the Sun) is similarly affected by GES remains lacking.}
  {This study aims to investigate, using a large sample of red giant stars, whether the Far-side exhibits signatures of the GES merger event, and to test whether the influence of GES on the Galactic inner halo is global through comparison with the Near-side.}
  {A sample of 14,056 red giant stars was selected from the SkyMapper Southern Survey Fourth Data Release (SMSS DR4) and the Gaia Third Data Release (Gaia DR3) through a series of stringent quality cuts designed to minimize disk contamination. We analyzed and compared the metallicity distributions, velocity-space distributions, velocity dispersions, anisotropy parameters, and action-space characteristics of the Near-side and the Far-side. We further demonstrate through a series of robustness tests that this global symmetry is not an artifact of measurement uncertainties, residual disk contamination, or incomplete removal of the Sagittarius stream.}
  {Observational evidence is found for the influence of the GES merger event on the Far-side. Chemical abundance analysis shows that the Far-side also contains a metal-rich component peaked at [Fe/H]~$\approx -1.4$. Kinematic analysis reveals that the metal-rich population ($-1.7 < $~[Fe/H]~$\leq -1.3$) on the Far-side exhibits GES-typical features consistent with the Near-side, namely a ``sausage''-shaped distribution in velocity space and a velocity anisotropy parameter as high as $\beta \sim 0.8$ -- 0.9. Action-space analysis further shows that $10.86\%\pm0.23\%$ of stars on the Far-side bear the dynamical imprint of the GES, statistically consistent with $11.56\%\pm0.16\%$ on the Near-side. The residual prograde rotation on the Far-side persists under increasingly aggressive disk-removal cuts, indicating that it reflects a genuine kinematic signal rather than residual disk contamination. These results demonstrate that the GES merger event affected not only the Near-side but also the global structure of the entire inner halo ($r < 28$~kpc), with its influence being present on the Far-side as well and exhibiting spatial symmetry.}
  {}
  

\titlerunning{GES influence on the opposite side of the Galactic Center}
\maketitle
\nolinenumbers

\section{Introduction}

Within the prevailing $\Lambda$CDM cosmological framework, galaxy formation is considered a process of hierarchical merging, where smaller systems are continuously accreted and fused to build massive galaxies like the Milky Way (MW) \citep{1978MNRAS.183..341W,1991ApJ...379...52W}. In this process, the stellar halo, as one of the oldest and most diffuse components in a galaxy, is widely regarded as a ``fossil record'' documenting these early merger events \citep{1999MNRAS.307..495H,1999Natur.402...53H,2018Natur.563...85H,2020A&A...642L..18K}). The long orbital periods of stars in the stellar halo allow accreted debris to maintain dynamical coherence over Gyr timescales, preserving imprints of their origin in both chemical abundances and kinematic patterns. Leveraging astrometry from the European Space Agency's \textit{Gaia} astrometric satellite \citep{2018A&A...616A...1G, 2021A&A...649A...1G} and chemical fingerprints from large-scale spectroscopic surveys, such as APOGEE \citep{2017AJ....154...94M}, LAMOST \citep{2012RAA....12.1197C}, and SDSS/SEGUE \citep{2009AJ....137.4377Y}, it has now become possible to unravel the assembly history of the MW stellar halo.

Among these, a milestone discovery was the confirmation that the MW experienced a violent major merger event at redshift $z \sim 2$ \citep[approximately 8 -- 11 billion years ago;][]{2018MNRAS.478..611B, 2018Natur.563...85H}. The remnants of this accreted dwarf galaxy were named the Gaia-Enceladus-Sausage (GES), which represents the last significant merger event of the MW. The GES progenitor was recognized as a massive dwarf galaxy, with stellar mass estimates ranging from \(10^{9} - 10^{10} M_{\odot}\) in earlier studies \citep{2019MNRAS.487L..47V, 2019MNRAS.490.3426D, 2019MNRAS.484.4471F, 2019MNRAS.482.3426M} to a more refined value of \(M_* \approx 1.45 \times 10^{8} M_{\odot}\) in recent work \citep{2023MNRAS.526.1209L}. A large amount of observational evidence indicates that the GES contributed a substantial portion of the stellar mass in the inner halo (typically considered $r < 30$ kpc), with estimates reaching as high as 50\% \citep{2021ApJ...923...92N}. The remnant stars from the GES possess distinct chemical and kinematic signatures: compared to other halo stellar components, they typically have higher metallicities, with a distribution peak located between [Fe/H] $\approx -1.1$ and $-1.5$ \citep{2020ApJ...901...48N, 2020MNRAS.492.3631M}, and exhibit a slight negative metallicity gradient, as first proposed by \citet{2022MNRAS.517.2787L}; simultaneously, their orbital motions exhibit extreme radial anisotropy, with the velocity-anisotropy parameter $\beta$ reaching as high as 0.9 \citep{2019AJ....157..104B, 2022MNRAS.517.2787L}. Recent studies have shown that the anisotropy ($\beta$) is strongly correlated with metallicity \citep{2019AJ....157..104B, 2021ApJ...919...66B}: metal-rich components exhibit extreme radial anisotropy ($\beta \sim 0.8$ -- $0.9$), while metal-poor components are relatively isotropic. This conclusion was further confirmed by \citet{2022MNRAS.517.2787L}, who also found, using RR Lyrae samples, that the inner halo ($r < 30$ kpc) is primarily dominated by metal-rich, high-radial-orbital components, which coincides with the dynamical remnant features of the GES fragments after merging. These stars on high-radial orbits present a typical ``sausage-like'' distribution in the velocity space 
($v_r$ -- $v_{\phi}$ plane), which is one of the most prominent observational features of the GES \citep{2018MNRAS.478..611B,Iorio2019,2021MNRAS.502.5686I}.

Most previous studies of the GES have either utilized all-sky samples for aggregate analyses without distinguishing between the near and far sides of the Galactic Center (GC; e.g., \citealp{2022MNRAS.517.2787L, 2019AJ....157..104B, 2021ApJ...919...66B,2022AJ....164...41W,2023ApJ...950..151W}), or have been limited by survey strategies to the Solar Neighborhood. Even all-sky studies with precise astrometry using 5-D information from an all-sky RR Lyrae sample \citep{2021MNRAS.502.5686I}, while providing a comprehensive picture of the inner halo, still rely on symmetry assumptions to reconstruct the full velocity ellipsoid and do not specifically probe the Far-side. \citet{2021ApJ...923...92N} also pointed out that their H3 survey sample was mainly based on North Galactic Cap data, further highlighting the observational bias toward specific regions. Currently, systematic detection of GES signals on the Far-side is lacking, in particular work conducted using red giants -- an important halo-population probe that is both abundant and capable of detecting further distances. Consequently, a systematic detection of GES signals on the Far-side, utilizing the full 6-D phase-space information and metallicity afforded by a large sample of red giants as tracers, remains a void that this study aims to fill.

This study uses a large sample of red giant stars selected from the SkyMapper Southern Survey Fourth Data Release \citep[SMSS DR4;][]{2024PASA...41...61O} and the Gaia Third Data Release \citep[Gaia DR3;][]{2023A&A...674A...1G}, providing full 6-D phase-space information and metallicity. We present the first systematic detection and comparative analysis of GES signals on both sides of the Sun-GC axis in the Galactic halo. The structure of this paper is arranged as follows: Section 2 introduces the data set and the sample selection and processing methods in detail; Section 3 presents our main analysis results, including the metallicity distribution, kinematic analysis, and dynamical analysis; Section 4 discusses the results; and Section 5 provides a summary.

\section{Data}
\label{sec:data}

\subsection{Sample selection}
\label{sec:sample}

This study uses the latest stellar catalog from the SPORTS (Stellar Parameters fOR all The Stars) project \citep{2025RNAAS...9...74H}. By integrating photometric data from SMSS DR4 and Gaia DR3, \cite{2025RNAAS...9...74H} provided physical parameters for 50 million stars. The derivation of metallicity ([Fe/H]), radial velocity, and distances in this catalog follows the core methodology described in \cite{2022ApJ...925..164H,2023ApJ...957...65H}.

Subsequently, using the observed parameters of these sample stars (celestial
coordinates, distance, proper motions, and radial velocity), we computed
their 3-D positions and 3-D velocities in the Galactocentric rest frame. We
adopted a right-handed Galactocentric Cartesian coordinate system
$(X, Y, Z)$, where the $X$-axis points in the direction opposite to the Sun,
the $Y$-axis points in the direction of Galactic rotation, and the $Z$-axis
points toward the North Galactic Pole. The velocities were also expressed in
a Galactocentric spherical coordinate system $(r, \theta, \phi)$, where $r$
is the Galactocentric distance, increasing radially outward, $\theta$ is the
polar angle, increasing toward the South Galactic Pole, and $\phi$ is the
azimuthal angle in the Galactic plane. We denote the corresponding velocity components by $v_r$, $v_\theta$ and
$v_\phi$, with $v_\phi$ measured along the direction of Galactic rotation,
so that $v_\phi > 0$ corresponds to prograde motion; with this convention
the Sun has $v_{\phi,\odot} = 247$~km~s$^{-1}$.
Throughout this paper, $\mu_{v_i}$, $\sigma_{v_i}$ and $\epsilon_{v_i}$
denote, respectively, the mean, the dispersion, and the per-star measurement
uncertainty of the velocity component $v_i$ ($i = r, \theta, \phi$), and
$\beta$ denotes the velocity anisotropy parameter, defined in
Sect.~\ref{sec:anisotropy}. In these calculations, we adopted the same
Galactic parameters as \citet{2022ApJ...925..164H}: the distance from the Sun
to the GC is set to $R_0 = 8.178$\,kpc \citep{2019A&A...625L..10G}; the Solar
peculiar velocity with respect to the Local Standard of Rest is
$(U_\odot, V_\odot, W_\odot) = (7.01, 10.13, 4.95)$\,km\,s$^{-1}$
\citep{2015MNRAS.449..162H}; and the total Solar azimuthal velocity is set to
$247$\,km\,s$^{-1}$ \citep{2009ApJ...705.1548R}.

The giant stars were extracted from this sample, totaling 13,378,604 stars. The following screening criteria based on stellar parameters were then applied. First, stars with valid radial velocity FLAG (where \texttt{FLAG\_RV} is not null) were retained, and the absolute radial velocity was limited to $|\texttt{rv}| \le 900$~km~s$^{-1}$. Secondly, the absolute Galactic latitude was restricted to $|b| \ge 10^\circ$, and the heliocentric distance was required to be positive ($\texttt{dist} > 0$). Finally, we require that the absolute values of the proper motions in both Right Ascension ($|\texttt{pmra}|$) and Declination ($|\texttt{pmdec}|$) be less than $999$\,mas yr$^{-1}$ to exclude measurement outliers. After applying these steps, 3,477,467 red giants remained.

To construct a Galactic halo sample, we applied stringent criteria to
minimize disk contamination and to ensure data quality. We selected stars
with a vertical distance from the Galactic plane $|Z| \geq 6$~kpc,
metallicity $\mathrm{[Fe/H]} < -1.3$, metallicity uncertainty
$\mathrm{err}_{\mathrm{[Fe/H]}} \leq 0.2$~dex, renormalized unit weight error
$\mathrm{RUWE} \leq 1.2$, and absolute spherical
velocity components $|v_r|, |v_\theta|, |v_\phi| < 500$~km~s$^{-1}$. These
cuts yielded 16,921 red giant stars.

To accurately investigate the properties of the GES, contamination from the
Sagittarius (Sgr) stream, one of the most prominent substructures in the
Galactic halo, must be minimized. We therefore removed potential Sgr stream
members in two steps. First, we transformed the equatorial coordinates
$(\alpha, \delta)$ of all stars into the heliocentric Sgr stream coordinates
$(\tilde{\Lambda}_\odot, \tilde{B}_\odot)$ using the rotation matrix of
\citet{2014MNRAS.437..116B}, in which $\tilde{B}_\odot = 0\degr$ is aligned
with the orbital plane of the stream, and retained as candidates only the
stars lying within $10\degr$ of that plane, i.e.
$|\tilde{B}_\odot| \leq 10\degr$. Second, among these candidates we
identified as Sgr members those whose heliocentric distances are consistent
with the distance track of the stream at their $\tilde{\Lambda}_\odot$,
following Equations~(2) and (3) of \citet{2019MNRAS.486..378L}, which are
based on the mean stream distance and line-of-sight depth tabulated by
\citet{2017ApJ...850...96H}. Using the distance information in addition to
the sky position avoids over-cleaning the sample, since many genuine halo
stars project close to the stream plane without belonging to it. Through
these steps, 2,865 Sgr stream members were excluded, resulting in a final
sample of 14,056 giant stars.

The final catalog includes the following parameters: Right Ascension
(\texttt{ra}), Declination (\texttt{dec}), Galactic longitude
(\texttt{gl}), Galactic latitude (\texttt{gb}), parallax
(\texttt{parallax}), proper motions in Right Ascension (\texttt{pmra}) and
Declination (\texttt{pmdec}), radial velocity (\texttt{rv}), metallicity
(\texttt{[Fe/H]}), and heliocentric distance (\texttt{dist}), together with
their associated observational uncertainties, all taken from the SPORTS
catalog \citep{2025RNAAS...9...74H}; the Galactocentric positions
$(X, Y, Z)$ and velocities $(U, V, W)$ in the Cartesian system; and the
velocities $(v_r, v_\theta, v_\phi)$ in the spherical system defined above.
We estimated the uncertainties of all the derived quantities via Monte Carlo
simulations: for each star, we resampled \texttt{dist}, \texttt{pmra},
\texttt{pmdec}, and \texttt{rv} 1000 times according to their observational
uncertainties, assuming Gaussian errors, propagated each set of resampled
observables through the full coordinate transformations, and adopted the
standard deviation of the 1000 realizations as the uncertainty of each
quantity. The resulting distance and velocity uncertainties are compared
between the two hemispheres in Sect.~\ref{sec:systematics}.

\subsection{Subsamples}
\label{sec:subsamples}
The final sample of 14,056 red giants was first divided into two subsamples based on their spatial distribution. A Galactocentric Cartesian coordinate system was adopted, where the $X=0$ plane passes through the GC and is perpendicular to both the Galactic disk and the Sun-GC line. In this system, the Sun is located at $X = - 8.178$ kpc, and the positive $X$-axis points from the Sun toward the GC.

\begin{figure*}[htbp]
   \sidecaption
   \includegraphics[width=0.62\textwidth, angle=0]{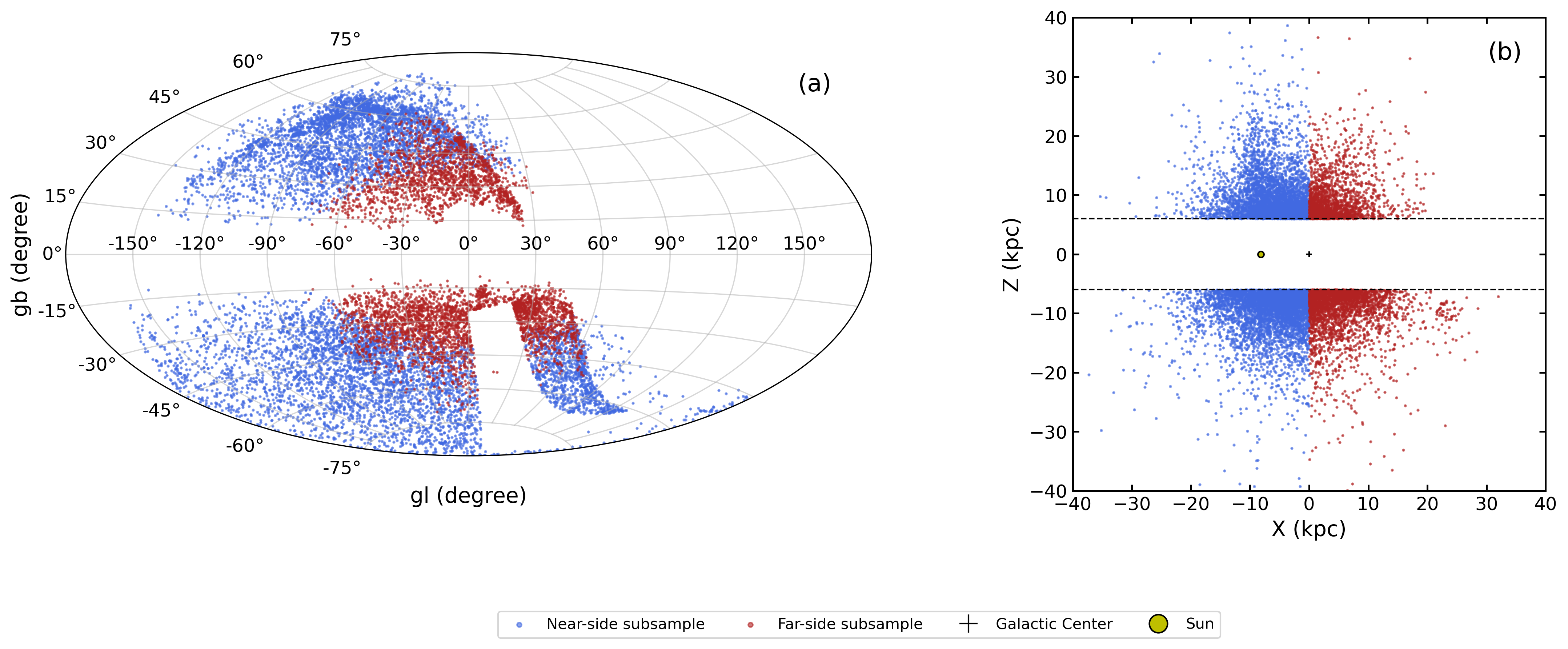}
   \caption{Spatial distribution of the stellar subsamples. Panel (a): All-sky Aitoff projection in Galactic coordinates ($gl$, $gb$). Panel (b): Distribution in the $X$--$Z$ plane of the Galactocentric Cartesian coordinate system. Blue and red symbols represent the N-side and F-side subsamples, respectively. The Sun is located at $X = -8.178$~kpc, and the GC is marked with a black cross. Horizontal dashed lines at $Z = \pm 6$~kpc denote the boundary used to exclude disk contaminants.}
   \label{fig:spatial_distribution}
\end{figure*}

According to this division, the following two subsamples were defined, with their spatial distribution illustrated in Figure~\ref{fig:spatial_distribution}:
\begin{itemize}
    \item Near-side (N-side) subsample: This subsample contains all stars with $X < 0$, comprising a total of 9,466 stars. It corresponds to the halo region on the same side of the GC as the Sun, which has been the focus of most previous studies.
    \item Far-side (F-side) subsample: This subsample contains all stars with $X > 0$, comprising a total of 4,590 stars. It corresponds to the halo region extending beyond the GC to the side opposite the Sun. This region is the main focus of this study.
\end{itemize}

Of the 2,865 Sgr stream members removed in Section~\ref{sec:data}, 
1,538 belong to the N-side and 1,327 to the F-side.

\begin{figure*}
\centering
\includegraphics[width=\textwidth]{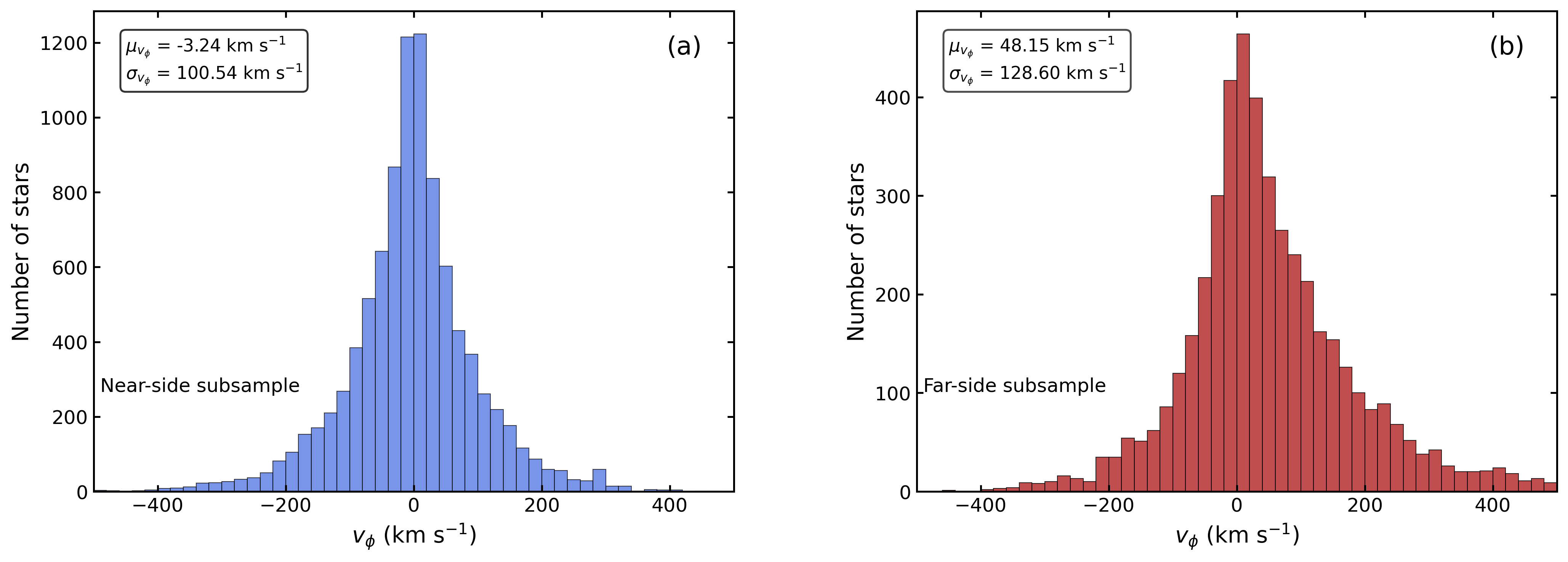}
\caption{Histogram of the $v_\phi$ distribution in Galactocentric spherical 
coordinates for the N- and F-side subsamples. Panel (a): The distribution of the N-side subsample, with 
$\mu_{v_\phi} = -3.24$~km~s$^{-1}$ and $\sigma_{v_\phi} = 100.54$~km~s$^{-1}$. 
Panel (b): The distribution of the F-side subsample, with 
$\mu_{v_\phi} = 48.15$~km~s$^{-1}$ and $\sigma_{v_\phi} = 128.60$~km~s$^{-1}$.}
\label{fig:vphi_strict}
\end{figure*}

Figure~\ref{fig:vphi_strict} presents the $v_\phi$ distributions of the 
two subsamples. The N-side subsample exhibits a symmetric, near-zero mean 
rotation of $\mu_{v_\phi} = -3.24$~km~s$^{-1}$ with a dispersion of 
$\sigma_{v_\phi} = 100.54$~km~s$^{-1}$, as expected for a pure halo 
population with little net rotation. In contrast, the F-side subsample 
shows a notably higher mean rotational velocity of 
$\mu_{v_\phi} = 48.15$~km~s$^{-1}$ with $\sigma_{v_\phi} = 128.60$~km~s$^{-1}$. 
We note that, relative to the large dispersion ($\sigma_{v_\phi} \approx 
100$~km~s$^{-1}$), the small mean rotation of the N-side is statistically 
consistent with zero and its sign carries no physical significance. The 
pronounced residual prograde signal on the F-side, however, is 
substantially larger and motivates a careful investigation into whether 
it reflects incomplete disk removal or a genuine physical asymmetry 
between the two hemispheres.

To distinguish between residual disk contamination and a genuine physical 
difference in the rotational properties of the two subsamples, we perform 
a series of robustness tests by applying progressively stricter quality 
cuts to the sample and examining whether the F-side mean rotational 
velocity $\mu_{v_\phi}$ persists. The rationale is straightforward: if the 
elevated F-side rotation were caused by residual disk stars, then 
strengthening the disk-removal cuts should progressively drive 
$\mu_{v_\phi}$ downward toward the N-side value, since disk stars are 
concentrated at low $|Z|$ and rotate rapidly. Conversely, if 
$\mu_{v_\phi}$ remains stable regardless of how aggressively the disk is 
removed, the signal cannot be attributed to disk contamination. The 
results are summarized in Table~\ref{tab:robustness}.

\begin{table*}
\centering
\caption{Robustness tests of the mean rotational velocity $\mu_{v_\phi}$ under progressively stricter vertical-height cuts.}
\label{tab:robustness}
\setlength{\tabcolsep}{8pt}
\begin{tabular}{lcccccc}
\hline\hline
Cut & $N_{\rm N}$ & $\mu_{v_\phi,\rm N}$ & $\sigma_{v_\phi,\rm N}$ & 
$N_{\rm F}$ & $\mu_{v_\phi,\rm F}$ & $\sigma_{v_\phi,\rm F}$ \\
 & & (km~s$^{-1}$) & (km~s$^{-1}$) & & (km~s$^{-1}$) & (km~s$^{-1}$) \\
\hline
No additional cut & 9466 & $-3.24$ & 100.54 & 4590 & 48.15 & 128.60 \\
$|Z| > 8$~kpc     & 5193 & $-3.84$ &  92.85 & 2470 & 38.21 & 117.25 \\
$|Z| > 10$~kpc    & 3081 & $ 1.36$ &  88.77 & 1416 & 37.23 & 111.17 \\
\hline
\end{tabular}
\tablefoot{$N$ denotes the number of stars retained after each cut.}
\end{table*}

We progressively raise the vertical-height threshold to $|Z| > 8$~kpc and 
$|Z| > 10$~kpc to exclude disk stars ever more aggressively. Under the 
$|Z| > 8$~kpc cut, the F-side $\mu_{v_\phi}$ is $38.21$~km~s$^{-1}$ 
(N-side: $-3.84$~km~s$^{-1}$); under the more stringent $|Z| > 10$~kpc cut, 
it remains $37.23$~km~s$^{-1}$ (N-side: $1.36$~km~s$^{-1}$), even as the 
samples shrink to 3,081 (N-side) and 1,416 (F-side) stars. Thus, although 
the disk-removal cut is tightened well beyond what is needed to eliminate 
the disk population, the F-side $\mu_{v_\phi}$ decreases only marginally, 
from $48.15$ to $\sim$37~km~s$^{-1}$, and never collapses toward the 
near-zero N-side value, while the N-side remains within a few km~s$^{-1}$ 
of zero throughout. The robustness of this asymmetry against increasingly 
aggressive disk-removal cuts strongly indicates that the elevated mean 
rotation on the F-side is not an artifact of disk contamination, but rather 
reflects a genuine kinematic signal. We note that, since $r \geq |Z|$ by definition, imposing $|Z| > 10$~kpc
leaves the $3 < r < 10$~kpc region empty, so this row of
Table~\ref{tab:robustness} probes only the $r > 10$~kpc halo.

\begin{figure}[htbp]
   \centering
   \includegraphics[width=0.48\textwidth]{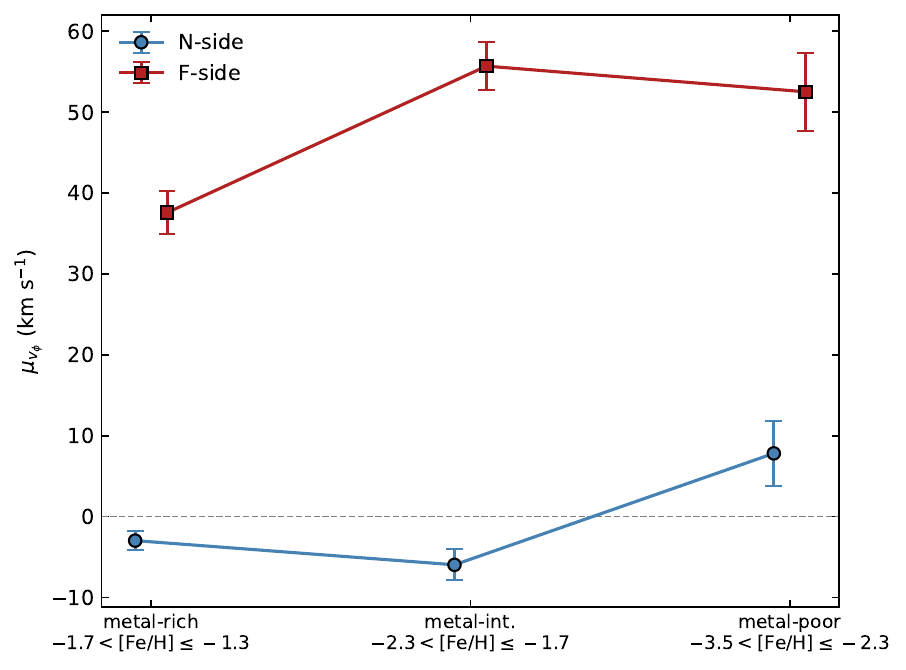}
   \caption{Mean rotational velocity $\mu_{v_\phi}$ as a function of
   metallicity bin for the N-side (blue circles) and F-side (red squares).
   Error bars denote the $1\sigma$ uncertainty from $1000$
   bootstrap resamples, and the gray dashed line marks zero rotation. The
   N-side rotation remains small in all bins, whereas the F-side shows a
   substantial prograde rotation in every metallicity bin,
   with the metal-rich bin somewhat lower, so the prograde signal
   is present across the full metallicity range rather than being confined
   to the metal-rich population.}
   \label{fig:muvphi_feh}
\end{figure}

To clarify the nature of this elevated F-side rotation, we examined its
dependence on metallicity, dividing each hemisphere into three metallicity
bins, namely the metal-rich ($-1.7 < \mathrm{[Fe/H]} \leq -1.3$),
metal-intermediate ($-2.3 < \mathrm{[Fe/H]} \leq -1.7$), and metal-poor
($-3.5 < \mathrm{[Fe/H]} \leq -2.3$) bins. Figure~\ref{fig:muvphi_feh}
shows the mean rotational velocity $\mu_{v_\phi}$ of the two hemispheres in
these bins. On the N-side, the rotation remains small in every bin, between
$-6.0$ and $7.8$~km~s$^{-1}$. On the F-side, by contrast, a substantial
prograde rotation is present in all three bins, with values of $37.6$,
$55.7$, and $52.5$~km~s$^{-1}$ from the metal-rich to the metal-poor bin,
and is in fact somewhat lower in the metal-rich bin. The prograde
signal is therefore present across the full metallicity range rather than
being confined to the metal-rich population, as would be the case if it
arose solely from a heated-disk or Splash component, which is typically
metal-rich \citep{2020MNRAS.494.3880B}. We further find that, under a $|Z|>10$~kpc
cut, the metal-rich F-side rotation decreases from $37.6$ to
$14.0$~km~s$^{-1}$, while the intermediate- and metal-poor values remain
essentially unchanged. This indicates that a Splash-like, disk-heated
component does contribute to the metal-rich rotation at low $|Z|$, but
cannot account for the prograde signal seen across all metallicities at
large $|Z|$.

This behavior points to a combination of mechanisms rather than a single
origin. The metal-rich, low-$|Z|$ component is consistent with an in-situ
or Splash population heated by the merger \citep{2020MNRAS.494.3880B}, whereas the
prograde rotation that persists across all metallicities and to large $|Z|$
is more naturally associated with the GES debris itself, which is known to
carry a mild prograde rotation \citep{2018Natur.563...85H}, and may be further
enhanced by reflex motion of the Sun or by a tilted velocity ellipsoid
projecting into $v_\phi$ along the F-side line of sight. A full
decomposition of these contributions is beyond the scope of the present
work. We note, however, that the origin of this rotation does not affect the
symmetry analysis presented below, which relies on the shape of the velocity
ellipsoid rather than on its net rotation.

\section{Analysis and results}
\label{sec:analysis}
In this section, the chemical, kinematic, and dynamic properties of the N-side and F-side are compared, to investigate evidence of the GES's impact on the opposite side of the GC.

\subsection{Metallicity distribution}
\label{sec:metallicity_dist}

\begin{figure*}[htbp]
   \sidecaption
   \includegraphics[width=0.62\textwidth, angle=0]{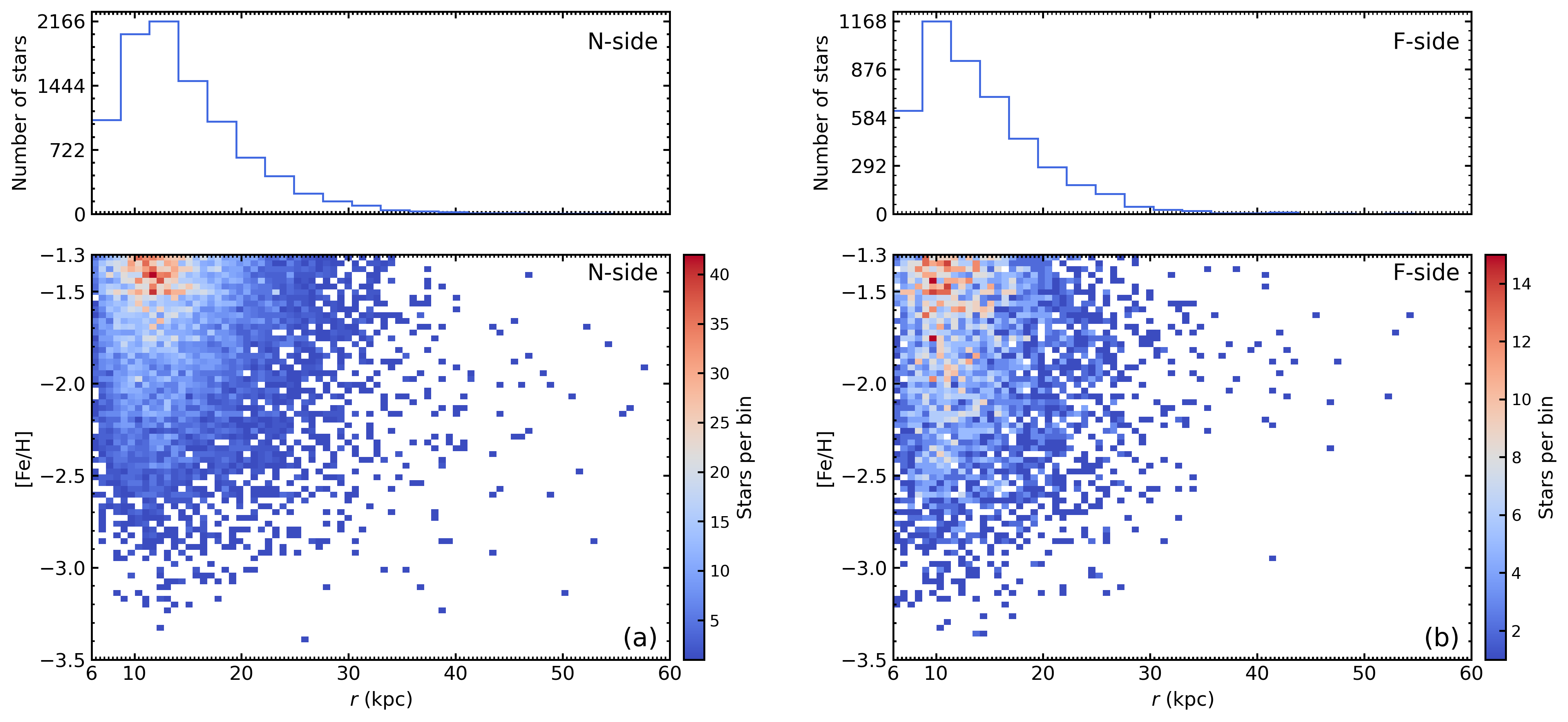} 
   \caption{Distribution density map of red giant stars in the Galactic halo in the plane of $\mathrm{[Fe/H]}$ and $r$. Panels (a) and (b) correspond to N-side and F-side samples, respectively. The main panels are 2-D density histograms, where the color represents the number of stars within each bin, the horizontal axis is the $r$, and the vertical axis is the $\mathrm{[Fe/H]}$. The top subplots display the 1-D radial number distribution histograms for the corresponding samples.}
   \label{fig:N_F_side_feh_r}
\end{figure*}

Figure \ref{fig:N_F_side_feh_r} displays the stellar number density distribution in the $r$ versus [Fe/H] plane for the N- and F-sides. The color in the figure represents the number of stars within each bin, while the histograms at the top illustrate the radial distribution characteristics of the samples. Both samples exhibit consistent distribution morphologies. As shown in Figure \ref{fig:N_F_side_feh_r}, the high-density core regions (dark red patches) for both samples are concentrated within $r < 15$ kpc. The histograms corroborate this, showing that star counts peak at $12$ kpc and subsequently decrease outward. The metallicity peak corresponding to this high-density core is located at $\mathrm{[Fe/H]} \approx -1.4$. This metal-rich component aligns well with the characteristics of the GES merger debris defined in the literature \citep{2018MNRAS.478..611B, 2018Natur.563...85H}, confirming that the inner halo region within $r < 28$ kpc is dominated by GES debris, regardless of whether it is on the N-side or F-side. This boundary of $r < 28$~kpc is adopted following \citet{2021ApJ...923...92N}, a radius that coincides with the well-established break in the stellar-halo density profile \citep{2011ApJ...731....4S, 2011MNRAS.416.2903D} and, in the H3 Survey, with the outer apocentric pile-up of the Gaia-Sausage/Enceladus debris \citep[double break at 12 and 28 kpc;][]{2022AJ....164..249H}, where it marks the apocenter pile-up of the GES debris in their fiducial merger model.

\subsection{Kinematics}
\label{sec:kinematics}

The N- and F-sides are considered in terms of their velocity distributions, velocity dispersions, and anisotropy parameters.

\subsubsection{Velocity distribution}
\label{sec:velocity_dist}

\begin{figure*}[htbp]
   \sidecaption
   \includegraphics[width=0.62\textwidth]{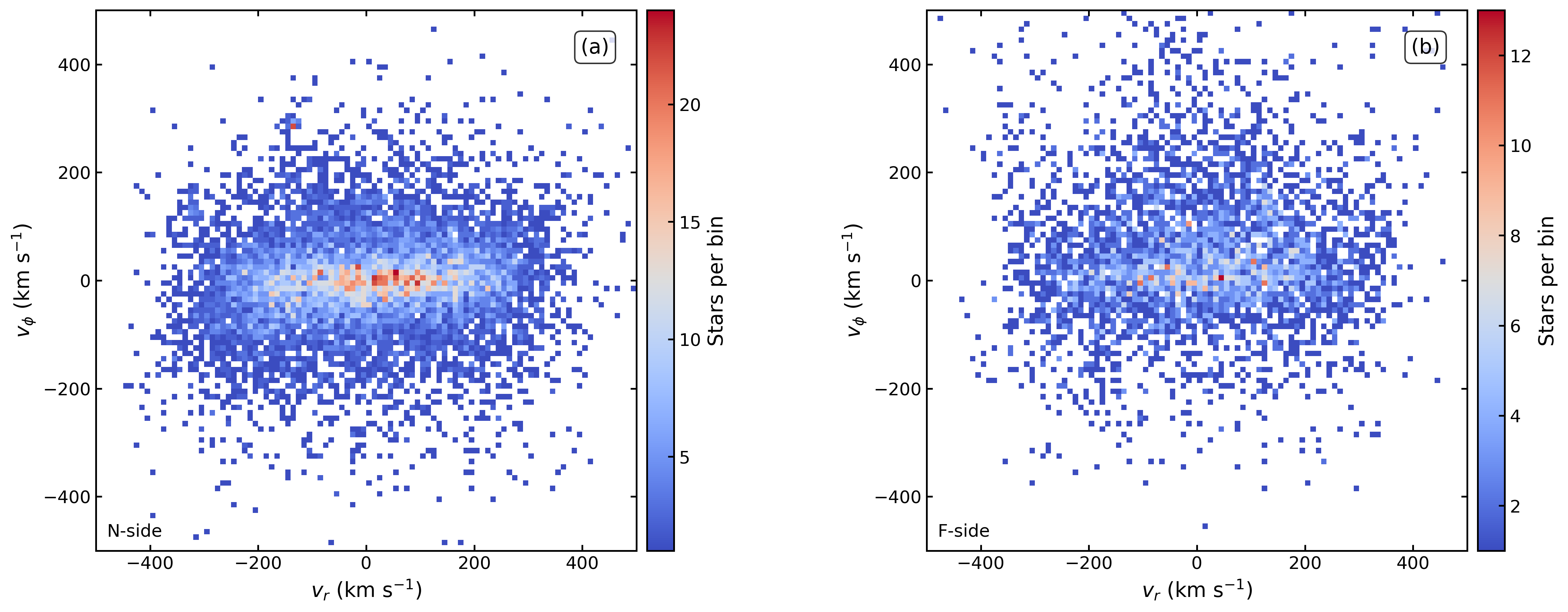}
   \caption{Distribution density map of the samples on both sides in the $v_r$--$v_\phi$ velocity plane. Panel (a) shows the N-side, and panel (b) shows the F-side. The color represents the number of stars within each velocity bin.}
   \label{fig:N_F_side_vr_vsphi}
\end{figure*}

The velocity distribution morphology of the N- and F-sides is first examined in the Galactocentric rest frame.
Figure \ref{fig:N_F_side_vr_vsphi} displays the 2-D distribution of $v_r$ and $v_{\phi}$ in spherical coordinates. The distributions are centered around $v_{\phi} \approx 0$ and elongated along the $v_r$ direction, exhibiting a prominent ``sausage-like'' structure. This result is consistent with the literature mentioned earlier in the Introduction.

\subsubsection{Velocity dispersion}
\label{sec:velocity_dispersion}

We now compare the velocity dispersions of the N- and F-sides. Each subsample
is split into three metallicity populations: metal-rich
($-1.7 < \mathrm{[Fe/H]} \le -1.3$), metal-intermediate
($-2.3 < \mathrm{[Fe/H]} \le -1.7$), and metal-poor
($-3.5 < \mathrm{[Fe/H]} \le -2.3$). For each population we compute
$\sigma_{v_r}$, $\sigma_{v_\theta}$ and $\sigma_{v_\phi}$ as a function of
$r$, in four radial bins: $3 < r < 10$, $10 < r < 15$, $15 < r < 20$, and
$20 < r < 60$~kpc. The widest outermost bin compensates for the decreasing
stellar density in the outer halo, thereby maintaining sufficient sample
statistics for the analysis. We estimated the dispersions using the
\texttt{Robust\_Sigma} algorithm \citep{1993ASPC...52..246L}, with
uncertainties derived from 1000 bootstrap resamples.  
To recover the intrinsic
dispersions, we subtracted the mean squared measurement uncertainty in each
bin in quadrature, $\sigma_{\rm true} = \sqrt{\sigma_{\rm obs}^{2} -
\langle\epsilon^{2}\rangle}$, where $\sigma_{\rm obs}$ is the dispersion
measured directly from the observed velocities and $\langle\epsilon^{2}\rangle$
is the mean of $\epsilon_{v_i}^{2}$ over the stars in that bin. This correction
follows \citet{2023MNRAS.525.5915Z}. All velocity dispersions and anisotropy
parameters reported in this work are based on these error-corrected values.

\begin{figure*}[htbp]
   \centering
   \includegraphics[width=0.8\textwidth, angle=0]{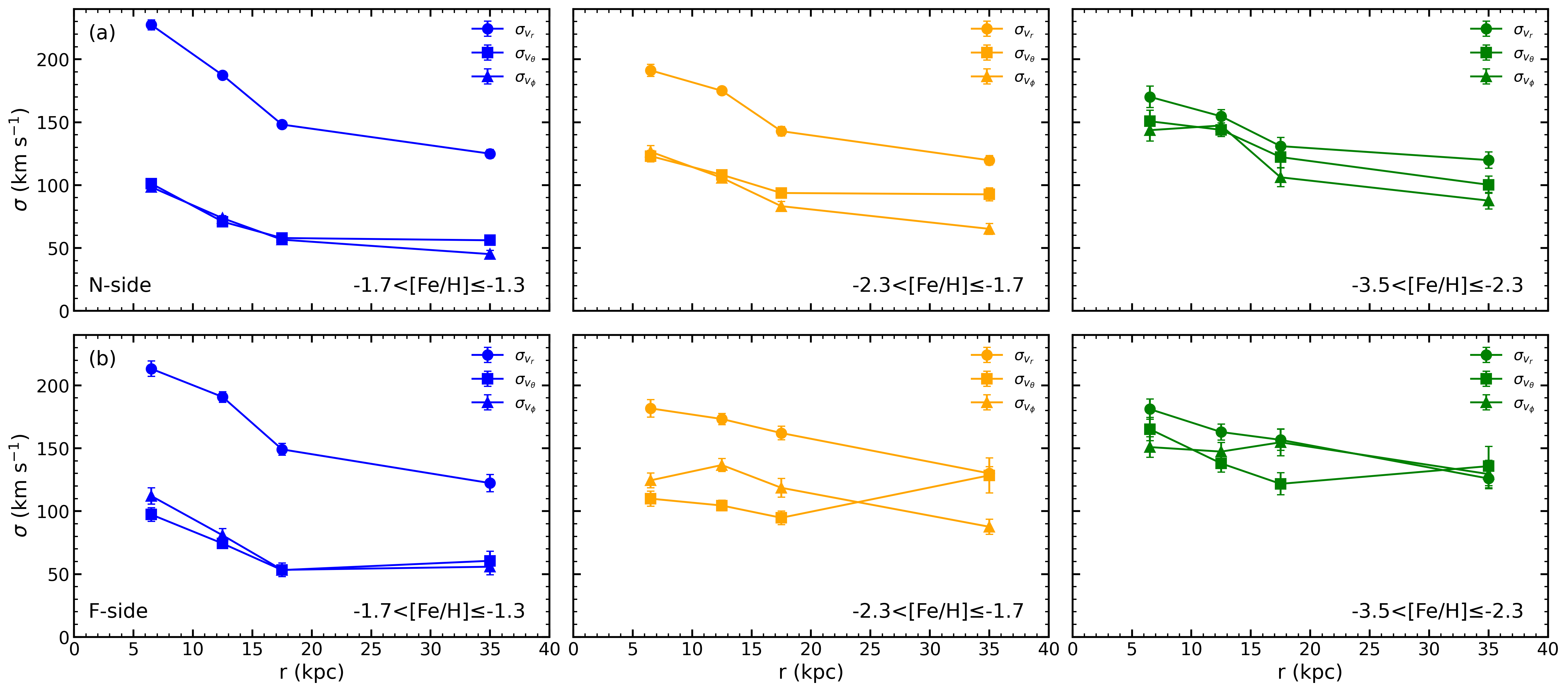}
    \caption{Velocity dispersion components $\sigma_{v_r}$ (circles),
   $\sigma_{v_\theta}$ (squares), and $\sigma_{v_\phi}$ (triangles) as a
   function of Galactocentric radius $r$. Top row (a): N-side; bottom row (b): F-side. Columns from left to right show the metal-rich
   ($-1.7 < \mathrm{[Fe/H]} \le -1.3$, blue), metal-intermediate
   ($-2.3 < \mathrm{[Fe/H]} \le -1.7$, orange), and metal-poor
   ($-3.5 < \mathrm{[Fe/H]} \le -2.3$, green) populations. For clarity, the $x$-axis is displayed out to 40~kpc, which fully covers the radial position of the outermost data point. The number of stars in each metallicity--radius bin is listed in Table~\ref{tab:nbin_sigma}.}
   \label{fig:N_F_side_sigma}
\end{figure*}

Figure~\ref{fig:N_F_side_sigma} presents the results. The metal-rich
populations (left panels) on both sides exhibit strong radially dominated
motion. Throughout the radial range, $\sigma_{v_r}$ is significantly higher
than $\sigma_{v_\theta}$ and $\sigma_{v_\phi}$; in the inner halo in
particular, $\sigma_{v_r}$ reaches approximately $180$ to $200$~km~s$^{-1}$,
while $\sigma_{v_\phi}$ remains around $70$ to $100$~km~s$^{-1}$. This
difference forms a prominent ``sausage-like'' structure in velocity space,
the typical kinematic imprint left by the GES merger event.

As metallicity decreases, the radially dominant trend gradually weakens. The metal-intermediate population shows transitional features, while the metal-poor population exhibits isotropy ($\sigma_{v_r} \approx \sigma_{v_\theta} \approx \sigma_{v_\phi}$). This kinematic trend with metallicity is highly consistent with the results observed in the large-scale stellar halo by \citet{2021ApJ...919...66B} using K giants and  blue horizontal-branch stars, and by \citet{2022MNRAS.517.2787L} using RR Lyrae stars. These results indicate that the kinematic structure of the F-side is consistent with the features of the N-side, namely, the inner halo is dominated by a metal-rich accreted component (GES) with strong radial anisotropy.

\begin{table*}[htbp]
\centering
\caption{Number of stars entering each velocity-dispersion bin of Fig.~\ref{fig:N_F_side_sigma}, for the three metallicity populations and four radial bins on the N-side and F-side.}
\label{tab:nbin_sigma}
\setlength{\tabcolsep}{6pt}
\begin{tabular}{lcccc}
\hline\hline
[Fe/H] bin & $3<r<10$ & $10<r<15$ & $15<r<20$ & $20<r<60$ \\
\hline
\multicolumn{5}{c}{N-side} \\
\hline
$-1.7<\mathrm{[Fe/H]}\le-1.3$ & 1003 & 2116 & 1093 & 718 \\
$-2.3<\mathrm{[Fe/H]}\le-1.7$ &  773 & 1299 &  811 & 563 \\
$-3.5<\mathrm{[Fe/H]}\le-2.3$ &  194 &  422 &  213 & 237 \\
\hline
\multicolumn{5}{c}{F-side} \\
\hline
$-1.7<\mathrm{[Fe/H]}\le-1.3$ &  481 &  711 &  364 & 186 \\
$-2.3<\mathrm{[Fe/H]}\le-1.7$ &  448 &  732 &  402 & 328 \\
$-3.5<\mathrm{[Fe/H]}\le-2.3$ &  244 &  364 &  194 & 133 \\
\hline
\end{tabular}
\tablefoot{As described in Sect.~\ref{sec:velocity_dispersion}, the analysis uses four radial bins spanning $3 < r < 60$~kpc, the outermost bin being kept wide to compensate for the declining stellar density of the outer halo. The tabulated counts sum to $9{,}442$ (N-side) and $4{,}587$ (F-side), marginally below the full subsample sizes ($9{,}466$ and $4{,}590$); the small differences (24 and 3 stars, respectively) are stars lying beyond $r = 60$~kpc.}
\end{table*}

\subsubsection{Anisotropy parameter}
\label{sec:anisotropy}
\begin{figure*}[htbp]
   \sidecaption
   \includegraphics[width=0.62\textwidth]{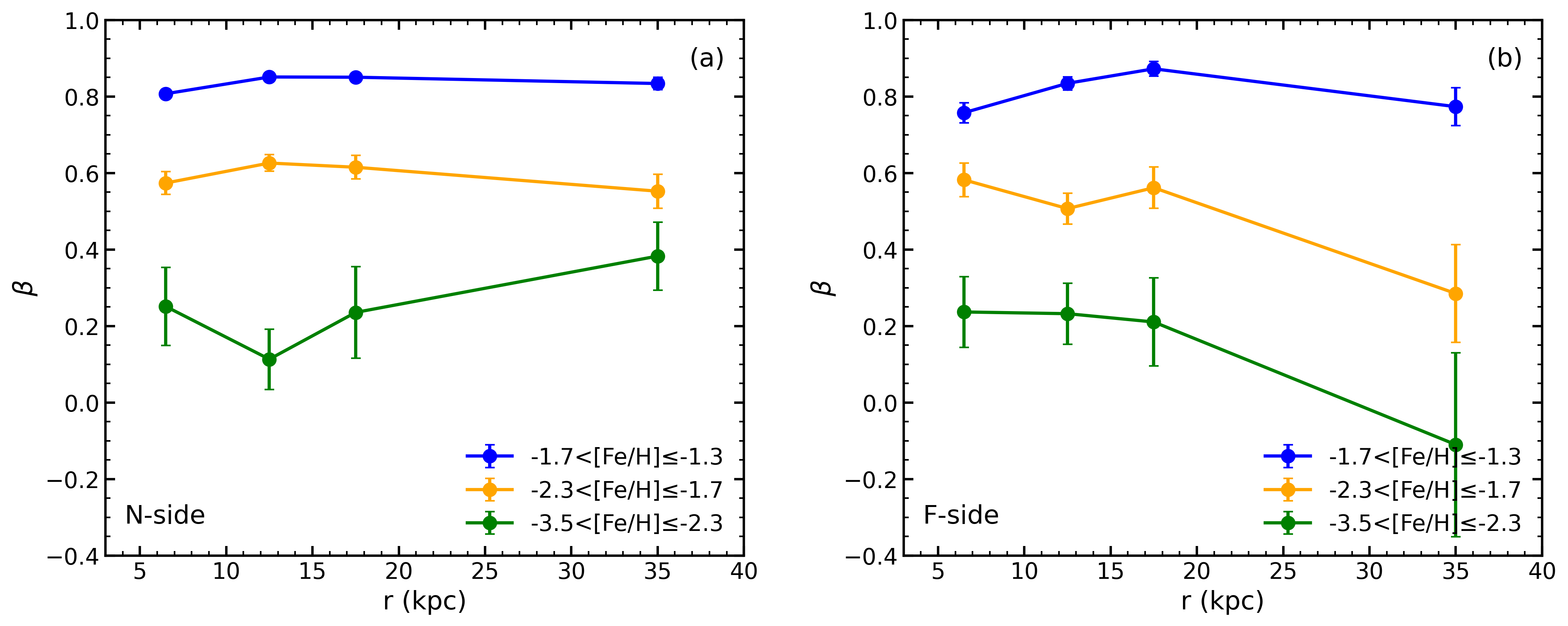}
   \caption{Panels (a) and (b) show the velocity anisotropy parameter $\beta$ 
   as a function of Galactocentric radius $r$ for three different metallicity 
   populations in the N-side and F-side samples, respectively.}
   \label{fig:N_F_side_beta}
\end{figure*}

To further investigate the kinematic properties of both sides, the velocity 
anisotropy parameter $\beta$ is calculated, defined as 
$\beta = 1 - (\sigma_{v_\theta}^2 + \sigma_{v_\phi}^2) / (2\sigma_{v_r}^2)$, 
based on the velocity dispersions calculated above. This parameter 
characterizes the shape of the velocity ellipsoid, indicating whether it is 
radially dominated, tangentially dominated, or isotropic. Figure 
\ref{fig:N_F_side_beta} displays the profiles of $\beta$ as a function of 
$r$ for different metallicity populations on both sides, revealing a strong 
correlation between metallicity and kinematic anisotropy.

The metal-rich populations on both sides (blue curves) exhibit extremely high 
radial anisotropy, maintaining a high $\beta$ level of around $0.8$ within 
the inner-halo range ($r < 28$ kpc). This extreme radial feature matches the 
$\beta \sim 0.9$ measured by \citet{2022MNRAS.517.2787L} using RR Lyrae stars 
in the inner halo. This confirms that the metal-rich population on the F-side 
is indeed GES, and is the radial orbit remnant it retained on the far side of 
the Galaxy.

As the metallicity decreases, the value of $\beta$ decreases significantly. 
The metal-poor populations on both sides (green curves) exhibit isotropy or 
even a slight tangential bias ($\beta \sim 0$ or slightly less than 0). This 
further reinforces the picture that the inner halo is composed of a 
metal-rich, radially anisotropic accreted component (GES) and a relatively 
metal-poor, isotropic background component. This analysis shows that the 
kinematic imprint of GES is clear and robust on both the N-side and the 
F-side. This two-component picture is in excellent agreement with the findings 
of \citet{2021MNRAS.502.5686I}, who demonstrated that the inner-halo RR Lyrae 
population can be decomposed into a metal-rich, radially anisotropic component 
and a metal-poor, isotropic component.

The remarkable consistency of $\beta \approx 0.8$--$0.9$ for the metal-rich 
population across both the N-side and F-side provides the strongest 
observational evidence for the global spatial symmetry of the GES remnants. 
This kinematic symmetry, manifested in the velocity ellipsoid shape, reflects 
the fact that GES debris occupies similarly radial orbits on both sides of the 
Galactic Center, demonstrating that the GES merger event shaped the inner halo 
globally rather than locally.

We have verified that this result is insensitive to the adopted value
of $R_0$; the test is presented in Appendix~\ref{app:r0}.
We have further verified that this radial anisotropy is robust against the
measurement uncertainties and the choice of sample, as discussed in
Section~\ref{sec:systematics}.

\subsection{Action space}

The action-space coordinates, $\mathbf{J} \equiv (J_R, J_{\phi}, J_Z)$, 
provide a natural and powerful framework for analyzing the dynamics of 
Galactic halo stars. Since actions remain adiabatically invariant during 
the slow evolution of the gravitational potential, this space is particularly 
suitable for identifying substructures and debris remnants generated by 
accretion events \citep{1984MNRAS.206..159B,2008gady.book.....B}. The 
distribution of stars is derived in action space, where $J_{\phi}$ 
corresponds to the angular momentum about the Galactic rotation axis, reflecting the rotational properties of the orbit; $J_R$ 
and $J_Z$ quantify the oscillation amplitudes in the radial and vertical 
directions, respectively.

Orbit integration and action calculation for the sample stars were performed 
using the Python Galactic dynamics library \texttt{galpy} 
\citep{2015ApJS..216...29B}. During the calculation, the 
\texttt{MWPotential2014} model \citep{2015ApJS..216...29B} was adopted to 
approximate the Galactic gravitational field. To eliminate the effect of 
Solar motion and normalize the parameters, the Solar kinematic parameters 
and normalization constants determined by \citet{2019MNRAS.484.2166S} 
(i.e., $J_{\phi,\odot}=2009.92~\mathrm{kpc~km~s^{-1}}$, 
$J_{Z,\odot}=0.35~\mathrm{kpc~km~s^{-1}}$, 
$E_{\odot}=-64943.61~\mathrm{km^2~s^{-2}}$) were adopted.

\begin{figure*}[htbp]
   \sidecaption
   \includegraphics[width=0.62\textwidth]{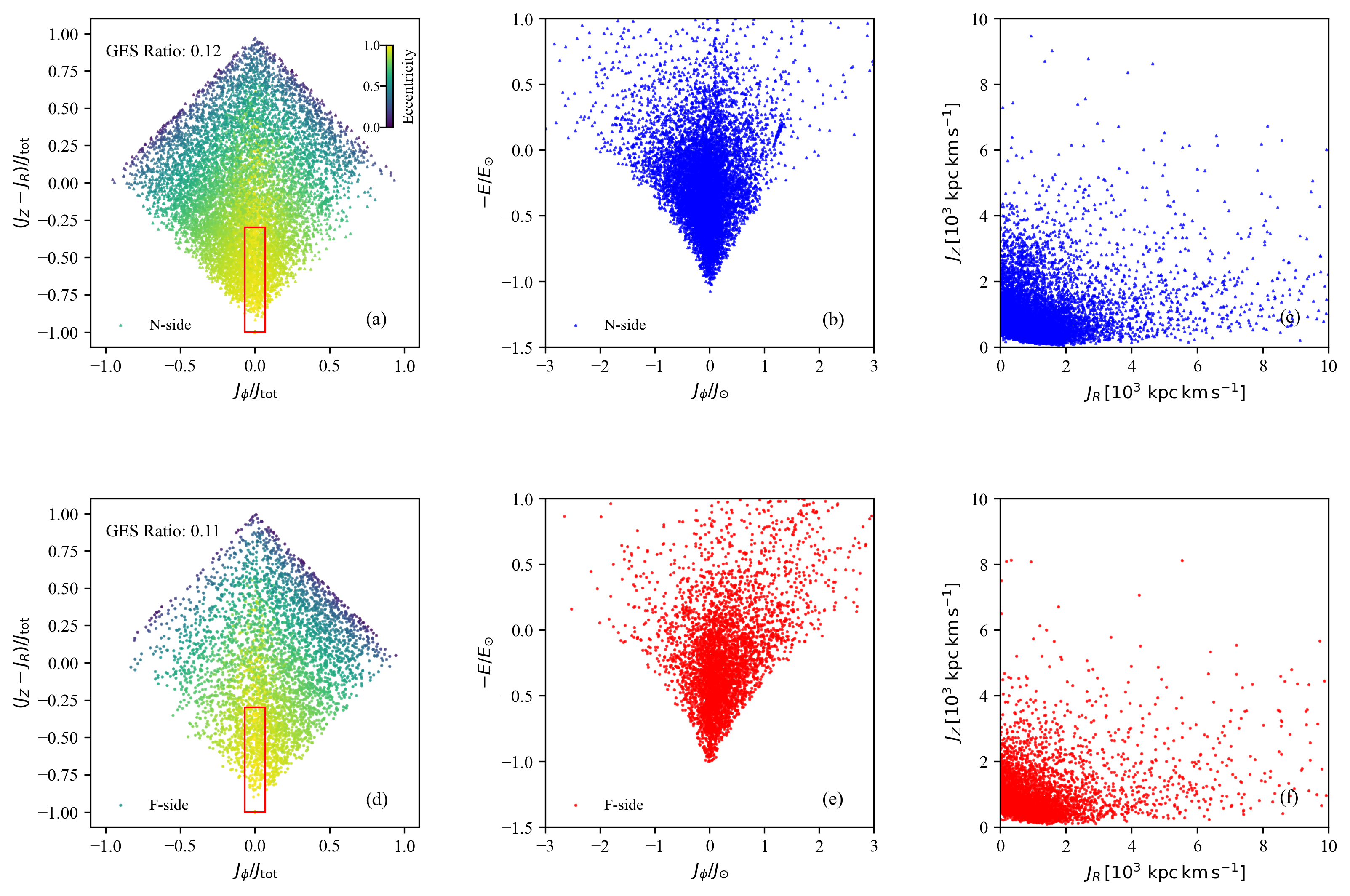}
   \caption{Dynamical property distributions of the halo red giant samples 
   on the N-side and F-side, respectively. Panels (a)--(c) show the results 
   for the N-side sample, while panels (d)--(f) show the results for the 
   F-side sample. (a) and (d) present the normalized action-space 
   distributions ($J_\phi/J_{\mathrm{tot}}$ versus $(J_Z - J_R)/J_{\mathrm{tot}}$), 
   where each dot represents an individual star. The color-coding indicates 
   the orbital eccentricity, with yellow representing high eccentricity and 
   purple representing low eccentricity. The red rectangles mark the dynamical 
   fingerprint region of the GES (defined following 
   \citealp{2019MNRAS.488.1235M}, i.e., $|J_{\phi}/J_{\mathrm{tot}}| < 0.07$ 
   and $(J_Z - J_R)/J_{\mathrm{tot}} < -0.3$), with the proportion of stars 
   falling into this region (GES Ratio) labeled in the top-left corner. 
   (b) and (e) display the distribution of normalized orbital energy ($E$) 
   as a function of azimuthal action ($J_\phi$). Note the characteristic 
   ``V''-shaped structure at $J_\phi \approx 0$. (c) and (f) show the 
   vertical action ($J_Z$) versus the radial action ($J_R$).}
   \label{fig:N_F_side_EJ}
\end{figure*}

The left panels (a) and (d) of Figure \ref{fig:N_F_side_EJ} present the 
relationship between the normalized azimuthal action $J_{\phi}/J_{\mathrm{tot}}$ 
and the difference between the vertical and radial actions 
$(J_Z - J_R)/J_{\mathrm{tot}}$. GES members are identified using the dynamical 
selection criteria proposed by \citet{2019MNRAS.488.1235M}, which requires 
stars to simultaneously satisfy two conditions: $|J_{\phi}/J_{\mathrm{tot}}| 
< 0.07$ and $(J_Z - J_R)/J_{\mathrm{tot}} < -0.3$. The former criterion 
selects stars with near-zero angular momentum, corresponding to highly radial 
orbits, while the latter criterion selects stars with dominant radial action 
over vertical action. Stars satisfying both criteria are indicated by the red 
rectangular box in the figure. The color coding represents the orbital 
eccentricity. For both the N- and F-side subsamples, a prominent population 
of high-eccentricity stars tightly clustered around zero angular momentum 
($J_{\phi} \approx 0$) is clearly observed.

This is a typical feature of radial anisotropy. The calculation shows that 
on the N-side, the fraction of stars falling into the GES fingerprint region 
is $11.56\%\pm0.16\%$, which is very close to the $10.86\%\pm0.23\%$ observed 
on the F-side, where the uncertainties are derived from 1000 bootstrap 
resamplings each drawing 80\% of the sample without replacement. A 
two-proportion $z$-test yields $z = 1.22$ and $p = 0.22$, indicating that 
the two fractions are statistically consistent within uncertainties. It should 
be noted that the selection criteria from \citet{2019MNRAS.488.1235M} are 
designed to isolate only the most prominent GES members, and the inferred 
fractions represent lower limits of the true GES contribution. We emphasize 
that this fraction consistency serves as a supplementary indicator, while the 
more robust evidence for global symmetry comes from the consistent $\beta$ 
profiles presented in Section~\ref{sec:anisotropy}.

The middle panels (b) and (e) of Figure \ref{fig:N_F_side_EJ} show the 
relationship between normalized energy $E$ and angular momentum $J_{\phi}$. 
Both samples exhibit the characteristic ``V-shape'' structure: The majority 
of stars are located at the bottom of the V (low angular momentum), while 
high-energy stars extend toward both sides. This further confirms that 
these stars are dominated by radial motion and include a fraction of 
high-energy retrograde members. The right panels (c) and (f) show the 
distribution of vertical action $J_Z$ versus radial action $J_R$. It can 
be seen that the samples exhibit extreme extensions along the $J_R$ axis 
(extending up to $10 \times 10^3~\mathrm{kpc~km~s^{-1}}$), indicating that 
these stars are on highly radial orbits \citep{2018ApJ...856L..26M}. This 
is consistent with the theoretical expectations for the GES as a nearly 
head-on collision merger event.

\section{Discussion}
\subsection{Comparison of results from the four-quadrant samples}
\label{sec:Comparison of results from the four-quadrant samples}

\begin{figure}[htbp]
   \centering
   \includegraphics[width=0.4\textwidth]{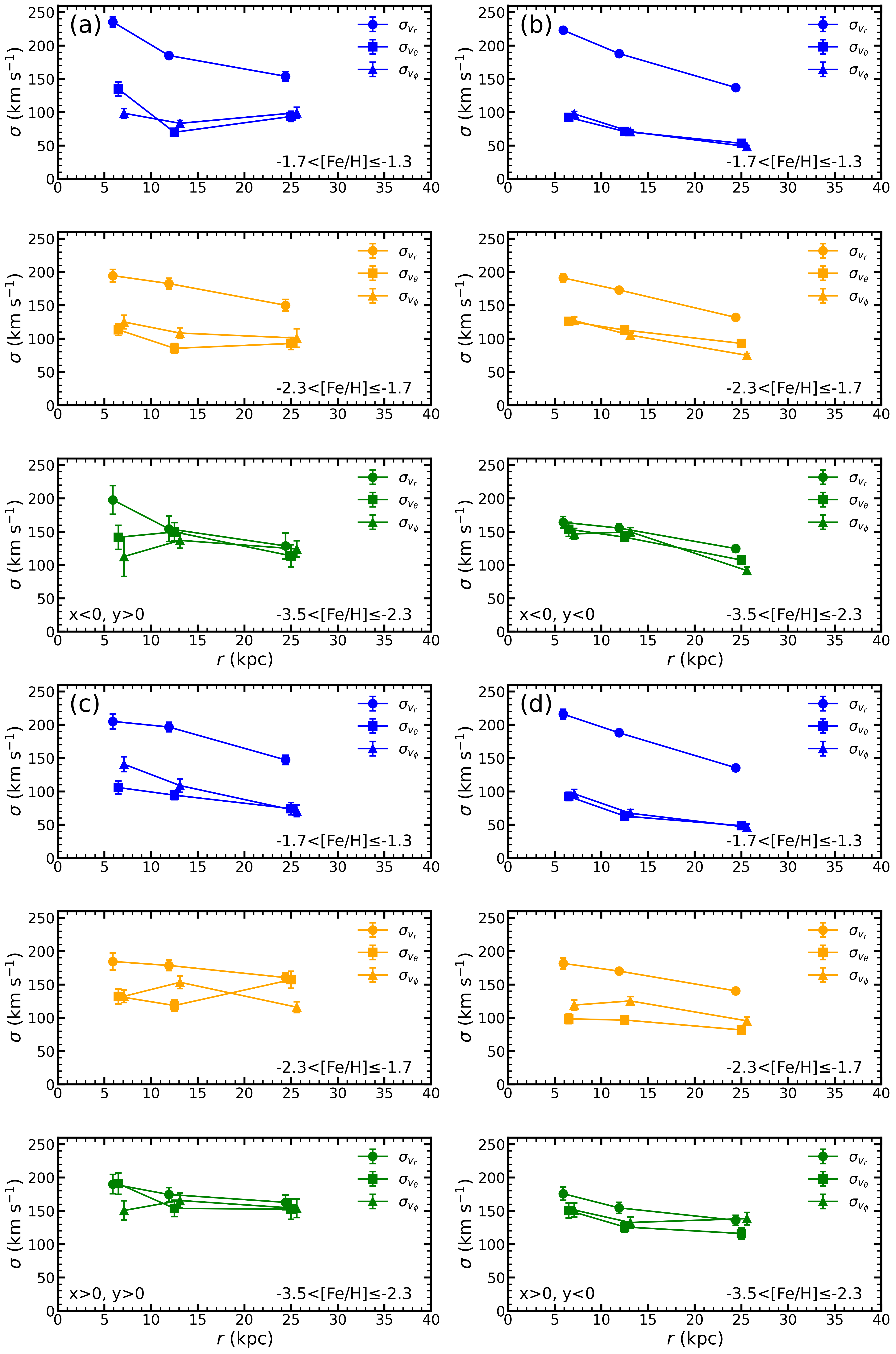}
   \caption{Velocity dispersion profiles ($\sigma_{v_r}$, $\sigma_{v_\theta}$,
   and $\sigma_{v_\phi}$) as a function of Galactocentric distance $r$ for the
   four spatial quadrants: (a) $x<0, y>0$, (b) $x<0, y<0$, (c) $x>0, y>0$, and
   (d) $x>0, y<0$. In each quadrant, the sample is subdivided into three
   metallicity bins: $-1.7 < \mathrm{[Fe/H]} \le -1.3$ (blue, top),
   $-2.3 < \mathrm{[Fe/H]} \le -1.7$ (orange, middle), and
   $-3.5 < \mathrm{[Fe/H]} \le -2.3$ (green, bottom). As in
   Fig.~\ref{fig:N_F_side_sigma}, circles represent $\sigma_{v_r}$, squares
   represent $\sigma_{v_\theta}$, and triangles represent $\sigma_{v_\phi}$.
   Error bars represent the $1\sigma$ standard deviation derived from 1000
   bootstrap resamples. In all four quadrants, the metal-rich component (top
   panels) exhibits a dominant radial dispersion
   ($\sigma_{v_r} > \sigma_{v_\theta}, \sigma_{v_\phi}$), representing the
   kinematic signature of the GES remnants. For clarity, the three components
   are slightly offset along the horizontal axis (this offset is for display
   purposes only and does not reflect a difference in radial position), and
   the $x$-axis is displayed out to 40~kpc, which fully covers the radial
   position of the outermost data point.}
   \label{fig:four_quadrants_sigma_c}
\end{figure}

\begin{figure}[htbp]
   \centering
   \includegraphics[width=0.4\textwidth]{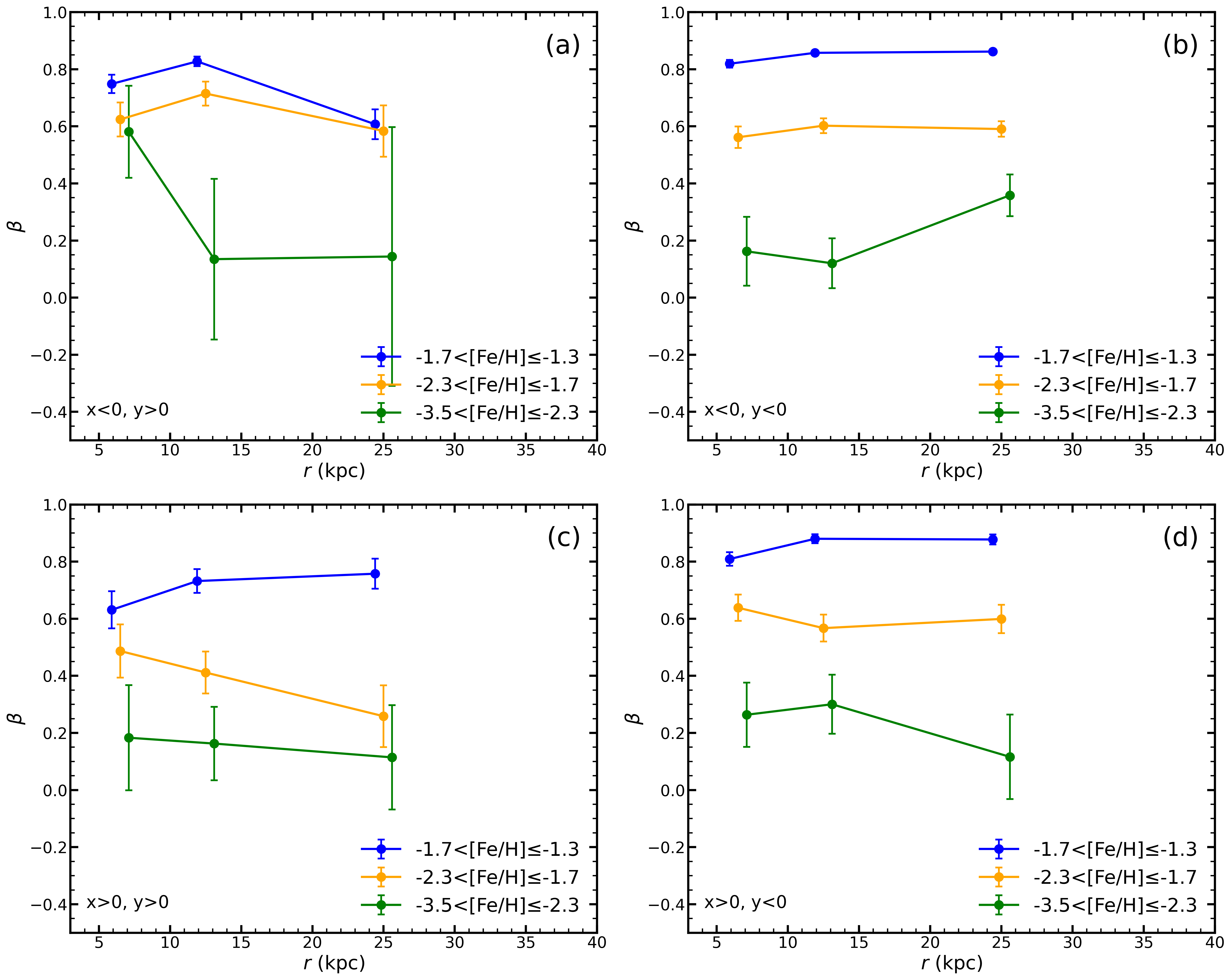}
   \caption{Velocity anisotropy parameter $\beta$ as a function of
   Galactocentric distance $r$ for the four spatial quadrants. Sub-panels
   (a)--(d) represent the same spatial regions as in
   Figure~\ref{fig:four_quadrants_sigma_c}. The three colors correspond to
   different metallicity ranges as defined in the legend. The metal-rich
   stars (blue) consistently exhibit high radial anisotropy, with $\beta$
   spanning $\sim 0.6$ to $0.9$ across all quadrants throughout the inner
   halo. As the metallicity decreases, the orbits transition toward a more
   isotropic state ($\beta \sim 0.1$ to $0.4$; the innermost metal-poor bin
   in the $x<0,\,y>0$ quadrant contains only 34 stars and is therefore
   poorly constrained, see Table~\ref{tab:nbin_quadrant}). The high degree
   of consistency across all quadrants confirms the global spatial symmetry
   and azimuthal invariance of the GES merger remnants. For clarity, points
   of different metallicity bins are slightly offset along the horizontal
   axis (this offset is for display purposes only and does not reflect a
   difference in radial position), and the $x$-axis is displayed out to
   40~kpc, which fully covers the radial position of the outermost data
   point.}
   \label{fig:four_quadrants_beta_c}
\end{figure}

To test whether the kinematic signatures of the GES are sensitive to
the azimuthal angle, we further subdivided the N- and F-side samples
into four quadrants along the $Y$-axis. Because the outer halo is
sparsely sampled once the data are split four ways, the two outermost
radial bins used in Sect.~\ref{sec:velocity_dispersion} were merged into
a single bin ($15 < r < 60$~kpc) for this analysis, so that each
quadrant retains sufficient statistics for a robust dispersion estimate;
the number of stars in every metallicity--radius bin is listed in
Table~\ref{tab:nbin_quadrant}. The velocity-dispersion and anisotropy
profiles are shown in Figs.~\ref{fig:four_quadrants_sigma_c} and
\ref{fig:four_quadrants_beta_c}, respectively.

As illustrated in Fig.~\ref{fig:four_quadrants_sigma_c}, for the metal-rich
component ($-1.7 < \mathrm{[Fe/H]} \le -1.3$), $\sigma_{v_r}$ markedly
exceeds $\sigma_{v_\theta}$ and $\sigma_{v_\phi}$ in all four quadrants,
reaching $\sim 205$ to $235$~km~s$^{-1}$ in the innermost radial bins. The
recurrence of this characteristic ``sausage-like'' velocity ellipsoid in
every quadrant provides compelling evidence that the GES debris does not
merely populate a preferred direction, but symmetrically fills the entire
inner-halo volume around the GC.

This azimuthal invariance is quantified by $\beta$
(Fig.~\ref{fig:four_quadrants_beta_c}). For the metal-rich population,
$\beta$ remains high and mutually consistent across all four quadrants,
spanning $\sim 0.6$ to $0.9$ throughout the inner halo. As the metallicity
decreases, $\beta$ declines monotonically toward isotropy in every quadrant,
with the metal-poor population ($-3.5 < \mathrm{[Fe/H]} \le -2.3$) typically
reaching $\beta \sim 0.1$ to $0.4$. This metallicity dependence of the
orbital anisotropy is therefore reproduced in all four azimuthal sectors. We
note that the metal-poor bins in the quadrant $x<0,\,y>0$ contain the fewest
stars (Table~\ref{tab:nbin_quadrant}) and consequently carry the largest
uncertainties. The innermost bin there contains only 34 stars and yields
$\beta = 0.58 \pm 0.16$, formally higher than in the other quadrants but
statistically poorly constrained. This does not affect the robust,
azimuthally invariant behavior of the metal-rich GES component, which is the
focus of this comparison. Taken together, the four-quadrant analysis
demonstrates that the extreme radial anisotropy of the metal-rich inner halo
is azimuthally uniform, confirming that the disrupted GES progenitor is now
distributed symmetrically on radial orbits spanning both sides of the GC.

\begin{table*}
\centering
\caption{Number of stars in each metallicity--radius bin of the four-quadrant analysis (Figs.~\ref{fig:four_quadrants_sigma_c} and \ref{fig:four_quadrants_beta_c}).}
\label{tab:nbin_quadrant}
\setlength{\tabcolsep}{6pt}
\begin{tabular}{llccc}
\hline\hline
Quadrant & [Fe/H] bin & $3<r<10$ & $10<r<15$ & $15<r<60$ \\
\hline
\multirow{3}{*}{$x<0,\,y>0$}
 & $-1.7<\mathrm{[Fe/H]}\le-1.3$ & 231 & 428 & 210 \\
 & $-2.3<\mathrm{[Fe/H]}\le-1.7$ & 170 & 235 & 129 \\
 & $-3.5<\mathrm{[Fe/H]}\le-2.3$ &  34 &  74 &  52 \\
\hline
\multirow{3}{*}{$x<0,\,y<0$}
 & $-1.7<\mathrm{[Fe/H]}\le-1.3$ & 772 & 1688 & 1601 \\
 & $-2.3<\mathrm{[Fe/H]}\le-1.7$ & 603 & 1064 & 1245 \\
 & $-3.5<\mathrm{[Fe/H]}\le-2.3$ & 160 &  348 &  398 \\
\hline
\multirow{3}{*}{$x>0,\,y>0$}
 & $-1.7<\mathrm{[Fe/H]}\le-1.3$ & 166 & 279 & 208 \\
 & $-2.3<\mathrm{[Fe/H]}\le-1.7$ & 154 & 281 & 295 \\
 & $-3.5<\mathrm{[Fe/H]}\le-2.3$ &  93 & 150 & 109 \\
\hline
\multirow{3}{*}{$x>0,\,y<0$}
 & $-1.7<\mathrm{[Fe/H]}\le-1.3$ & 315 & 432 & 342 \\
 & $-2.3<\mathrm{[Fe/H]}\le-1.7$ & 294 & 451 & 435 \\
 & $-3.5<\mathrm{[Fe/H]}\le-2.3$ & 151 & 214 & 218 \\
\hline
\end{tabular}
\tablefoot{The two outermost radial bins of Sect.~\ref{sec:velocity_dispersion} are merged into a single $15 < r < 60$~kpc bin to maintain sufficient statistics after the sample is split into four quadrants. As in Table~\ref{tab:nbin_sigma}, the tabulated counts exclude the small number of stars lying beyond $r = 60$~kpc.}
\end{table*}

\subsection{Global symmetry from the normalized velocity-distribution
difference}
\label{sec:diff_map}

The diagnostics presented above compare the two hemispheres through a set
of summary statistics ($\beta$ profiles, dispersion profiles, and GES
action-space fractions). As a complementary and largely model-independent
test, we now compare the full 2-D velocity
distributions of the two sides directly, without first decomposing them
into physical components. If the GES debris is genuinely symmetric across
the Galactic Center, the normalized velocity distributions of the N-
and F-side should be statistically indistinguishable in the region
occupied by the GES, and any coherent residual should be confined to the
populations that are known to differ between the two sides.

\begin{figure}[htbp]
   \centering
   \includegraphics[width=0.48\textwidth]{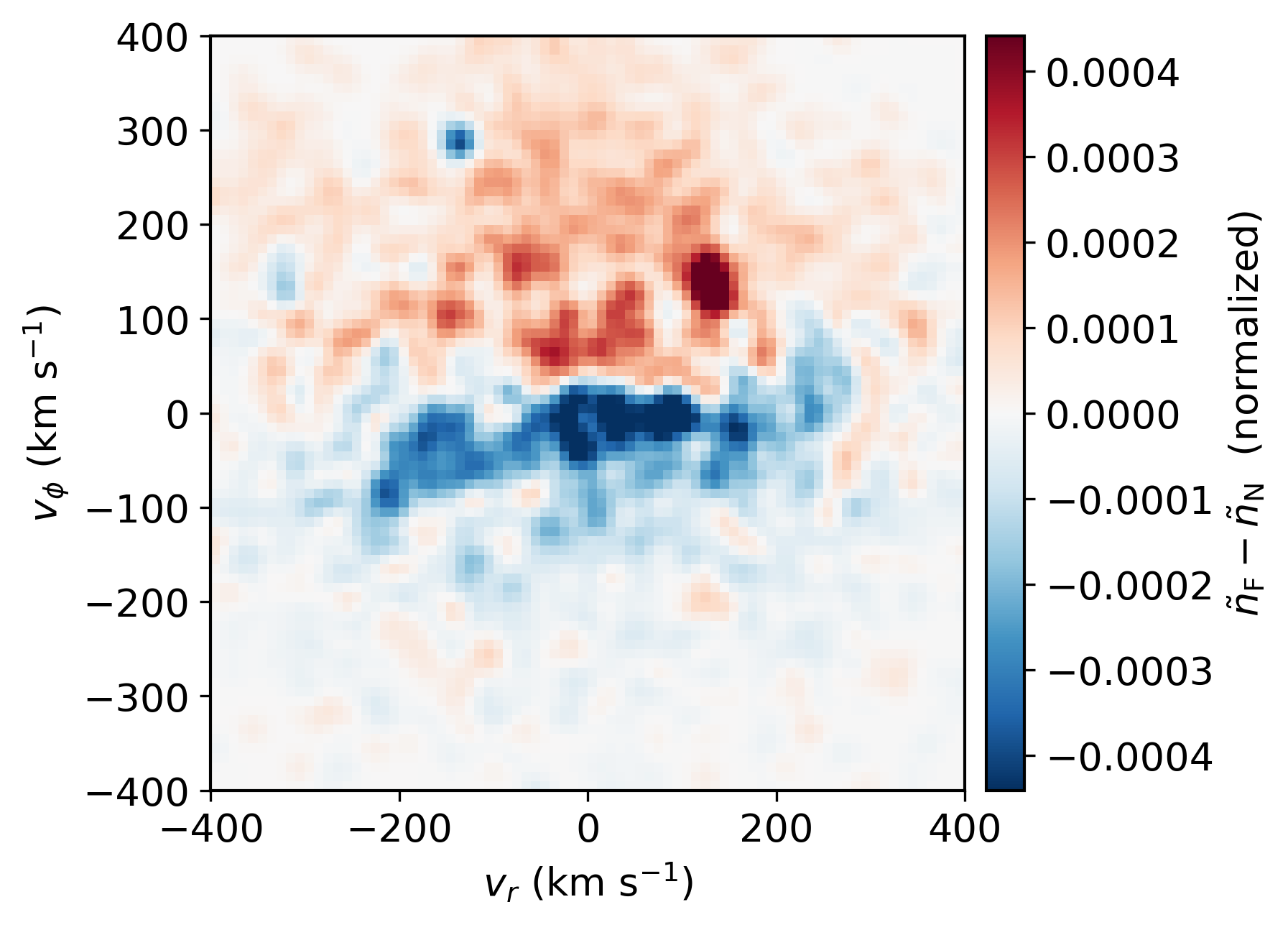}
   \caption{Difference between the normalized number densities of the
   F- and N-side subsamples in the $v_r$--$v_\phi$ plane
   ($\tilde{n}_{\rm F} - \tilde{n}_{\rm N}$, i.e. the F-side normalized
   density minus the N-side one), for the inner halo ($r < 28$~kpc).
   Here $\tilde{n}_{\rm F}$ and $\tilde{n}_{\rm N}$ are the normalized
   number densities of the F- and N-side, respectively. Both
   distributions were smoothed and normalized to unit sum before
   subtraction, so the map is independent of the different total star
   counts of the two sides. Red indicates a F-side excess (positive
   difference) and blue a N-side excess (negative difference). In the
   low-angular-momentum GES region ($|v_\phi| \lesssim 75$~km~s$^{-1}$)
   the two sides cancel to the noise level (median
   $|D/\sigma_D| \approx 1.1$, where $\sigma_D$ is the per-cell bootstrap
   uncertainty of $D$; stable over $|v_\phi| < 75$ to $150$~km~s$^{-1}$). The only coherent residual is the prograde
   F-side excess at $v_\phi \approx +100$ to $+200$~km~s$^{-1}$,
   corresponding to the residual prograde rotation of the F-side
   discussed in Section~\ref{sec:subsamples}. The global total variation
   distance between the two normalized distributions is
   $\mathrm{TVD} = 0.214$.}
   \label{fig:diff_map}
\end{figure}

We restrict the sample to the inner halo ($r < 28$~kpc), where the GES
dominates, and build 2-D number-density histograms of the two sides in the
$v_r$--$v_\phi$ plane on a common grid. Each histogram is lightly smoothed with a Gaussian kernel and normalized to unit sum, so that it represents a
probability density. We denote the normalized density of the N-side by
$\tilde{n}_{\rm N}$ and that of the F-side by $\tilde{n}_{\rm F}$, where
the subscripts N and F stand for the N- and F-side respectively, and
the tilde indicates that the density has been normalized. This
normalization makes the comparison insensitive to the different total star
counts of the two subsamples (9{,}091 on the N-side and 4{,}476 on the
F-side within $r<28$~kpc). We then form the difference map
$D \equiv \tilde{n}_{\rm F} - \tilde{n}_{\rm N}$, that is, in each grid
cell we subtract the N-side normalized density from the F-side one,
so that $D$ is the difference of the two normalized densities
(Figure~\ref{fig:diff_map}). The per-cell uncertainty $\sigma_D$ is
estimated from 1000 bootstrap resamplings of both subsamples and
represents the level of random fluctuation expected in $D$ at that cell;
the subscript distinguishes it from the velocity dispersions
$\sigma_{v_i}$ used elsewhere in this work. The ratio $D/\sigma_D$ is the
signal-to-noise map, giving the factor by which the difference in a cell
exceeds the noise: $|D/\sigma_D| \approx 1$ indicates a difference
comparable to the noise, whereas $|D/\sigma_D| \approx 3$ to $4$ indicates
a difference that is significant rather than a chance fluctuation. We use
this map to distinguish genuine residuals from Poisson noise.

Two features are apparent. First, in the GES sausage region, defined as
the low-angular-momentum band $|v_\phi| \lesssim 75$~km~s$^{-1}$ where the
radially elongated GES ridge dominates both subsamples, the two sides cancel to the noise level, with a median residual significance of
$|D/\sigma_D| \approx 1.1$, consistent with statistical fluctuations. The
prominent GES ridge therefore leaves no significant imprint in the
difference, indicating that its shape is the same on both sides. Second,
the only coherent, high-significance residual is a prograde feature at
$v_\phi \approx +100$ to $+200$~km~s$^{-1}$, where the F-side shows an
excess over the N-side. This feature lies outside the GES region and
coincides with the residual prograde rotation of the F-side identified
in Section~\ref{sec:subsamples}, rather than reflecting any asymmetry of
the GES debris itself. Because the normalization is global, this extra
prograde population slightly lowers the relative weight of the F-side
sausage, which accounts for the shallow complementary band of opposite
sign seen at low $v_\phi$.

We quantify the overall agreement with the total variation distance (TVD)
between the two normalized distributions, defined as
$\mathrm{TVD} = \tfrac{1}{2}\sum_i |\tilde{n}_{{\rm F},i} -
\tilde{n}_{{\rm N},i}|$, where the sum runs over all grid cells indexed by
$i$, and $\tilde{n}_{{\rm F},i}$ and $\tilde{n}_{{\rm N},i}$ are the
normalized densities of the F- and N-side in the $i$-th cell; that is, for
each cell we take the absolute difference of the two normalized densities,
sum over all cells, and divide by two. The TVD ranges from 0 for identical
distributions to 1 for no overlap, and requires no choice of any velocity
threshold. We obtain $\mathrm{TVD} = 0.214$. Because the TVD has no absolute
scale, with two finite samples drawn from a single underlying distribution
yielding a nonzero value, we calibrate it against a null distribution in
which the two hemispheres share identical velocity distributions. We
repeatedly draw from the N-side two subsamples with the actual N- and
F-side sizes and compute the TVD between their normalized maps. This null
distribution is tightly concentrated at $\mathrm{TVD}_{\rm null} = 0.099 \pm 0.004$ and never exceeds $0.115$ over
$1000$ realizations, which demonstrates that finite sampling alone produces a value of about $0.099$. The observed value of $0.214$ lies above this null
distribution, so the two hemispheres are not identical but differ by the
residual prograde rotation of the F-side already quantified in
Section~\ref{sec:subsamples}, while remaining globally similar in all other
respects. We further verified that the observed TVD is stable against the
analysis choices, varying only between $0.188$ and $0.266$ for grid
resolutions of $60$ to $100$ bins per axis and Gaussian smoothing scales of
$0.8$ to $1.6$ bins.
The interpretive inside and outside contrast is moreover robust to the
adopted definition of the GES region. Varying the threshold over
$|v_\phi| < 75, 100, 120,$ and $150$~km~s$^{-1}$ leaves the in-region
residual significance essentially unchanged, with $|D/\sigma_D| \approx 1.1$
to $1.3$, while the out-of-region relative residual grows monotonically from
$0.63$ to $0.90$, so the coherent difference always remains confined to the
prograde, out-of-GES region independent of the boundary. This statistic
requires no such calibration, since $|D/\sigma_D|$ already measures the
residual in units of its own noise. This direct and decomposition-free
comparison provides an independent confirmation that the GES kinematic
imprint is symmetric across the Galactic Center.

\subsection{Robustness against sample selection and measurement uncertainties}
\label{sec:systematics}

The F-side subsample is smaller than the N-side one and may carry larger
measurement uncertainties. It is therefore important to assess whether the
symmetry observed between the two sides reflects a genuine physical
property of the GES remnant, or could instead be an artifact of the larger
uncertainties on the F-side.

\subsubsection{Quantification of the uncertainties on both sides}

\begin{figure}[htbp]
   \centering
   \includegraphics[width=0.48\textwidth]{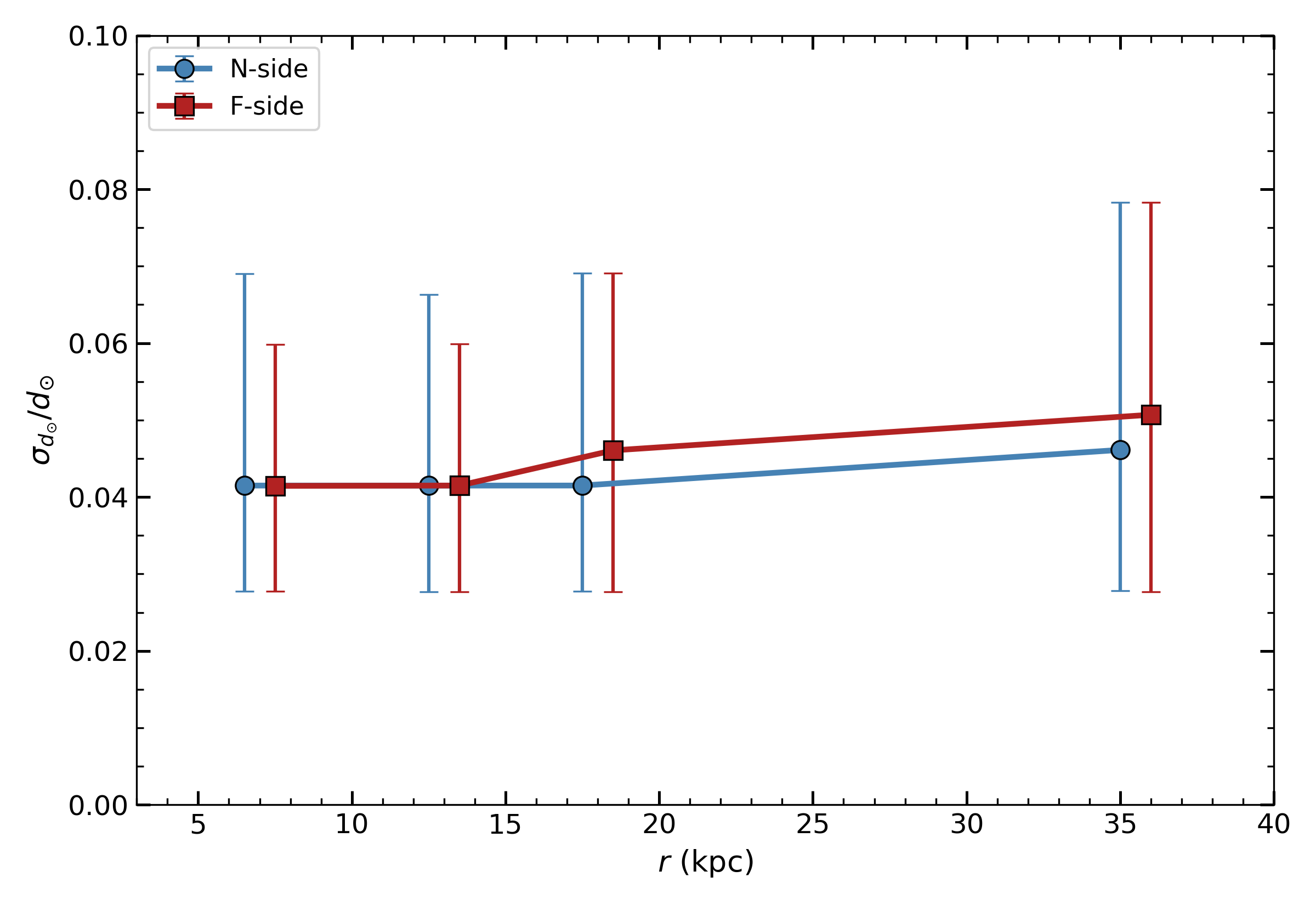}
    \caption{Median per-star relative heliocentric distance uncertainty $\sigma_{d_\odot}/d_\odot$ as a function of $r$ for the N-side (blue circles) and F-side (red squares). This quantity measures the random (formal) precision of the distances, i.e. the component $\sigma_{\rm dist,rand}$ of the distance error; it is $\approx 4$ to $5\%$ and nearly identical on the two sides. Combined in quadrature with the $\approx 16\%$ method accuracy
   $\sigma_{\rm dist,method}$ of the distance scale, the total per-star distance uncertainty is $\sigma_{\rm dis} = \sqrt{\sigma_{\rm dist,rand}^{2} + \sigma_{\rm dist,method}^{2}} \approx 17\%$. Error bars span the 16th to 84th percentiles, and the F-side points are offset by $+1$~kpc for clarity.}
   \label{fig:relerr_dist}
\end{figure}

\begin{figure*}[htbp]
   \centering
   \includegraphics[width=\textwidth]{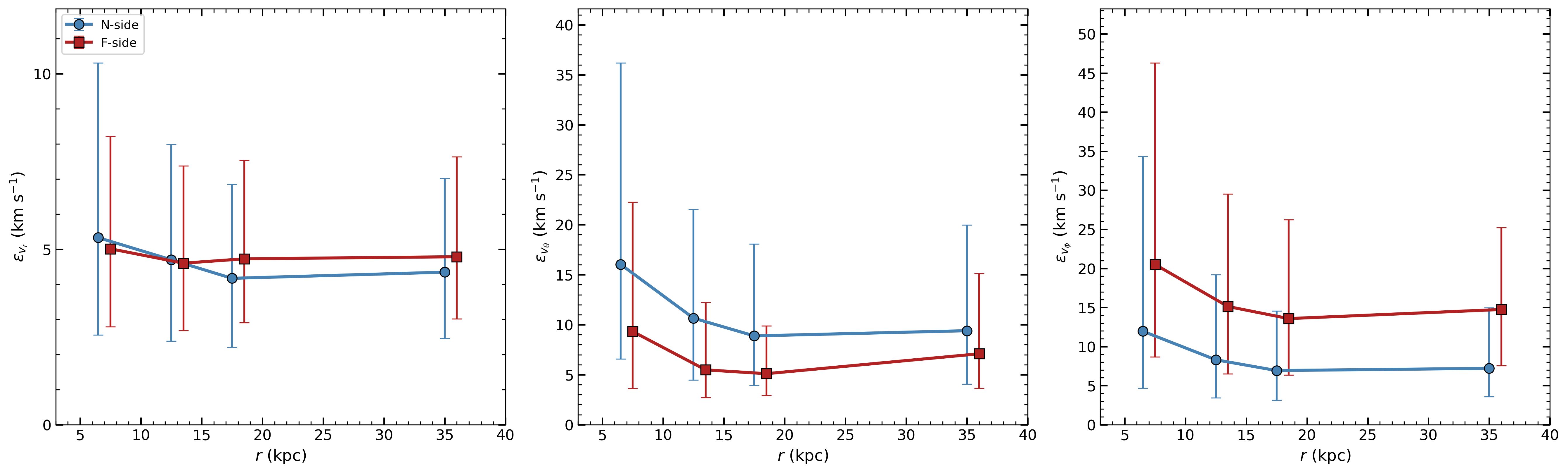}
   \caption{Median per-star measurement uncertainties of the three
   Galactocentric spherical velocity components, $\epsilon_{v_r}$ (left),
   $\epsilon_{v_\theta}$ (middle), and $\epsilon_{v_\phi}$ (right), as a
   function of $r$ for the N-side (blue) and F-side (red). These per-star
   uncertainties are distinct from the velocity dispersions $\sigma_{v_i}$ of
   Section~\ref{sec:velocity_dispersion} and are more than an order of
   magnitude smaller than the latter. Error bars span the 16th to 84th percentiles, and the F-side points
are offset by $+1$~kpc for clarity. The $x$-axis is displayed out to
40~kpc, which fully covers the radial position of the outermost data
point.}
   \label{fig:err_vel}
\end{figure*}

Using the Monte Carlo uncertainties described in Section~\ref{sec:data}
and the same radial bins as in Section~\ref{sec:velocity_dispersion}, we
compare the two sides in terms of the relative heliocentric distance
uncertainty $\sigma_{d_\odot}/d_\odot$ (Fig.~\ref{fig:relerr_dist}) and the
per-star uncertainties of the three Galactocentric spherical velocity
components, $\epsilon_{v_r}$, $\epsilon_{v_\theta}$, and $\epsilon_{v_\phi}$
(Fig.~\ref{fig:err_vel}). The median per-star relative distance uncertainty
$\sigma_{d_\odot}/d_\odot$ is nearly identical on the two sides ($\approx 4\%$
within $r < 15$~kpc, rising to $\approx 5\%$ in the outermost bins). This
quantity measures the random (formal) precision of the distances, which we
denote $\sigma_{\rm dist,rand}$, and does not include the accuracy of the
distance scale itself. The distance method of \citet{2022ApJ...925..164H},
which follows \citet{2019ApJS..243....7H}, was validated against
globular-cluster member stars, yielding a method (systematic) accuracy that we
denote $\sigma_{\rm dist,method} \approx 16\%$ \citep{2019ApJS..243....7H}. We
therefore adopt as the final per-star distance uncertainty the quadrature
combination of these two terms,
$\sigma_{\rm dis} = \sqrt{\sigma_{\rm dist,rand}^{2} + \sigma_{\rm dist,method}^{2}}$,
where $\sigma_{\rm dist,rand} \approx 4\%$ to $5\%$ is the random precision and
$\sigma_{\rm dist,method} \approx 16\%$ is the external method accuracy. This
gives a total distance uncertainty of $\sigma_{\rm dis} \approx 17\%$,
consistent with the $10$--$20\%$ range typical of photometric distances for
red giants and nearly identical on the two sides. Our conclusions do not depend
on the exact distance precision: the velocity dispersions are corrected for
measurement errors, this precision is essentially the same on the Near- and
Far-side so it cannot introduce a spurious N/F asymmetry, and the
error-injection experiment below confirms that the F-side's larger measurement
uncertainties do not create the signal.
$\epsilon_{v_r}$ is $\approx 4$ to $5$~km~s$^{-1}$ on both sides, while the
two tangential components are comparable in magnitude on the two sides, with
$\epsilon_{v_\theta}$ slightly larger on the N-side and $\epsilon_{v_\phi}$
slightly larger on the F-side (medians in the range $\sim 5$ to
$20$~km~s$^{-1}$). In all cases these uncertainties remain more than an order
of magnitude smaller than the inner-halo velocity dispersions
($\sigma_{v_r} \sim 150$ to $230$~km~s$^{-1}$). Moreover, the dispersions
reported in this work are already corrected for measurement errors
(Sect.~\ref{sec:velocity_dispersion}), so that the slightly larger tangential
uncertainty on the F-side does not bias the comparison.

\subsubsection{Error-injection experiment}

\begin{figure*}[htbp]
   \centering
   \includegraphics[width=\textwidth]{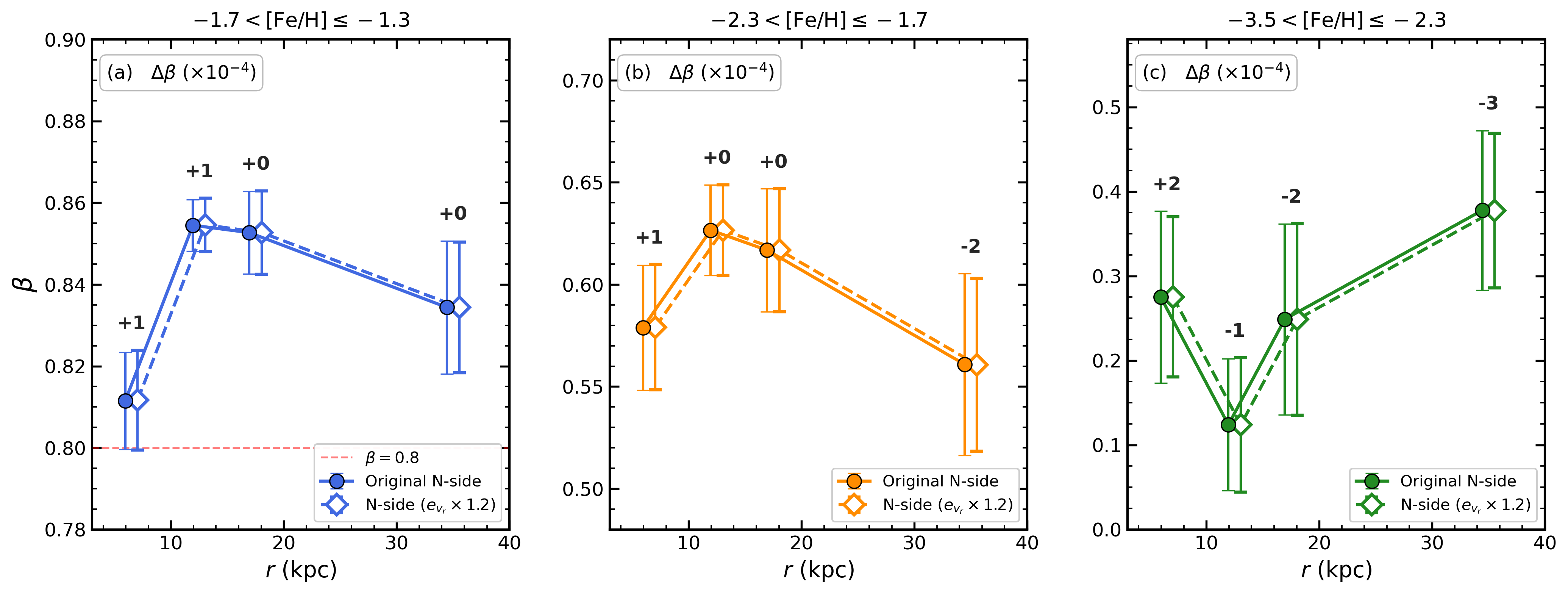}
   \caption{Error-injection test on the N-side sample. Panels (a) to (c) show
   $\beta$ as a function of $r$ for the metal-rich, metal-intermediate, and
   metal-poor populations, comparing the original N-side (filled circles,
   solid lines) with the N-side after inflating \texttt{err\_rv} by a factor
   of 1.2, the median F-side/N-side ratio (open diamonds, dashed lines). The
   number above each radial bin is
   $\Delta\beta = \beta(e_{v_r}\times1.2) - \beta(\mathrm{original})$, the
   change in $\beta$ produced by the injection, in units of $10^{-4}$. Note
   that the vertical scale differs between panels and is expanded relative to
   the full range of $\beta$, so that the error bars are clearly visible. In
   every bin $|\Delta\beta| \leq 3 \times 10^{-4}$, whereas the $1\sigma$
   bootstrap uncertainty on $\beta$ ranges from $0.006$ to $0.11$; the
   injection therefore shifts $\beta$ by far less than its statistical
   uncertainty, and the metal-rich $\beta$ remains at $\sim 0.85$. Error bars
   are $1\sigma$ from 1000 bootstrap resamples. For clarity the two sets of
   points are slightly offset along the horizontal axis; this offset is for
   display purposes only and does not reflect a difference in radial position.
   The red dashed line in panel (a) marks $\beta = 0.8$.}
   \label{fig:inj_beta}
\end{figure*}

To test most directly whether the F-side signal could be produced by its
larger uncertainties, we performed an error-injection experiment on the
N-side, whose GES signal is well established. The median radial-velocity
uncertainty is systematically larger on the F-side than on the N-side
(medians of $\sim 4.5$ to $4.8$ vs. $\sim 3.7$ to $4.2$~km~s$^{-1}$ across
the radial bins, i.e. a ratio of $\approx 1.2$). We therefore inflated the
N-side \texttt{err\_rv} by a factor of 1.2 and re-propagated it through the
same Monte Carlo procedure to obtain an N-side sample with F-side-level
measurement uncertainties, for which $\sigma_{v_i}$ and $\beta$ were
recomputed exactly as in Sect.~\ref{sec:kinematics}. As shown in
Fig.~\ref{fig:inj_beta}, the anisotropy parameter of the sample with inflated
uncertainties is almost indistinguishable from the original one across all
three metallicity bins. Quantitatively, the change produced by the injection
never exceeds $|\Delta\beta| = 3 \times 10^{-4}$ in any metallicity--radius
bin, whereas the $1\sigma$ bootstrap uncertainty on $\beta$ itself ranges from
$0.006$ to $0.11$. The injection therefore shifts $\beta$ by far less than its
statistical uncertainty, and the metal-rich $\beta$ remains at $\sim 0.85$.
Injecting F-side-level uncertainties into the N-side thus neither creates nor
distorts the GES kinematic signature, so the comparable uncertainties
intrinsic to the F-side cannot be responsible for its GES signal. This is
consistent with the statistical nature of the errors: the observed dispersion
$\sigma_{\rm obs}$, the intrinsic dispersion $\sigma_{\rm true}$, and the
per-star measurement uncertainty $\epsilon$ are related by
$\sigma_{\rm obs}^{2} = \sigma_{\rm true}^{2} + \epsilon^{2}$, so measurement
errors inflate the observed dispersion. Because the tangential dispersions
($\sigma_{v_\theta}$, $\sigma_{v_\phi}$) are intrinsically smaller than the
radial one ($\sigma_{v_r}$), an error of the same magnitude inflates them
proportionally more, which biases
$\beta = 1 - (\sigma_{v_\theta}^{2}+\sigma_{v_\phi}^{2})/(2\sigma_{v_r}^{2})$
toward isotropy. Measurement errors can thus only weaken, never create, the
radial anisotropy.

\subsubsection{Extinction and bulge/bar contamination}

Finally, the geometric cuts $|b| \ge 10^\circ$ and $|Z| \ge 6$~kpc ensure
that the lines of sight to the F-side halo stars lie several kpc away from
the Galactic plane, largely avoiding the bulge/bar region and the
high-extinction zone near the midplane. The concern that extinction and
crowding could simultaneously bias the photometric metallicities and the
astrometry therefore has limited impact on our sample. Taken together,
these tests demonstrate that the N-side/F-side symmetry is a genuine
physical property of the GES remnant rather than a consequence of sample
selection or measurement uncertainties.

\subsection{Completeness of Sagittarius stream removal}
\label{sec:sgr_completeness}

Residual contamination from the Sgr stream could in principle affect the
chemical and kinematic analysis presented in this work. In this section we
therefore assess the completeness of the Sgr stream removal described in
Sect.~\ref{sec:data} and its potential impact on our conclusions.

\begin{figure*}[htbp]
   \centering
   \includegraphics[width=\textwidth]{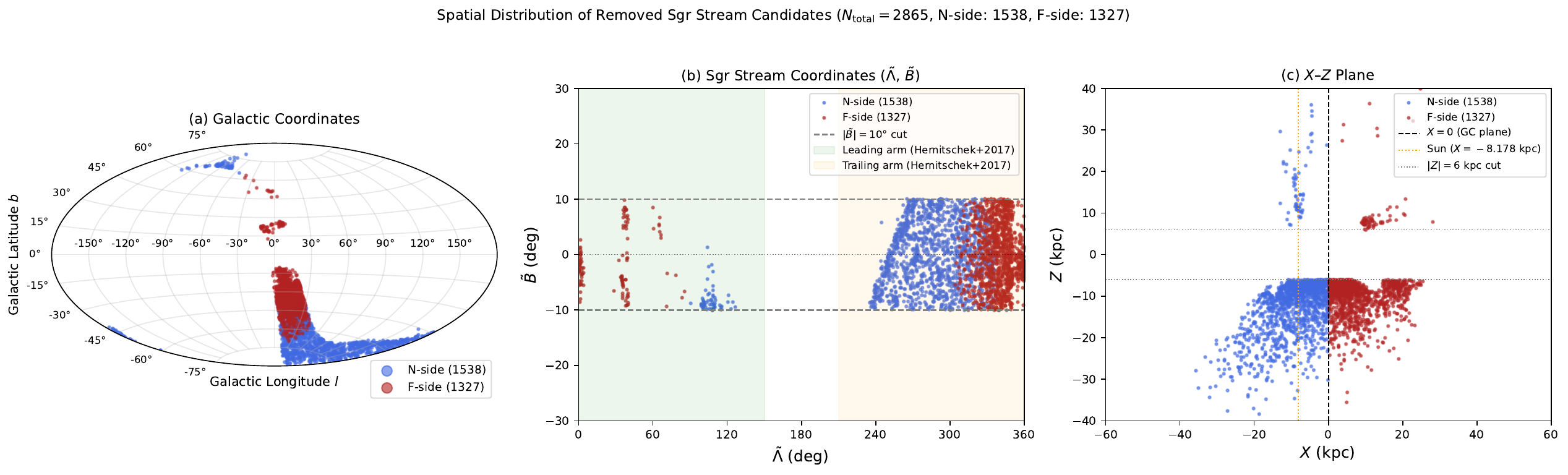}
   \caption{Spatial distribution of the 2,865 Sagittarius (Sgr) 
   stream members removed from our sample 
   (N-side: 1,538; F-side: 1,327). 
   Panel (a): All-sky Aitoff projection in Galactic coordinates 
   ($l$, $b$). 
   Panel (b): Distribution in the heliocentric Sgr stream coordinates 
   $(\tilde{\Lambda}_\odot, \tilde{B}_\odot)$ of \citet{2014MNRAS.437..116B}. 
   The gray dashed lines mark the $|\tilde{B}_\odot| = 10\degr$ boundary used 
   to define Sgr stream candidates. The green and orange shaded regions 
   indicate the angular extent of the leading and trailing 
   arms as defined by \citet{2017ApJ...850...96H}, respectively. 
   Panel (c): Distribution in the $X$--$Z$ plane of the 
   Galactocentric Cartesian coordinate system. 
   Blue and red symbols represent the N-side and F-side 
   subsamples, respectively. 
   The removed stars trace the expected arc-like morphology 
   of the Sgr stream, confirming that the removal criteria 
   accurately identify genuine Sgr members.}
   \label{fig:sgr_spatial}
\end{figure*}

Figure~\ref{fig:sgr_spatial} shows the spatial distribution of the 
2,865 removed Sgr stream candidates in Galactic coordinates, Sgr 
stream coordinates $(\tilde{\Lambda}_\odot, \tilde{B}_\odot)$, and the $X$--$Z$ 
plane. The removed stars trace the expected arc-like morphology of 
the Sgr stream, with the majority concentrated in the trailing arm 
region ($210\degr \leq \tilde{\Lambda}_\odot \leq 360\degr$), confirming 
that the removal criteria accurately identify genuine Sgr members 
rather than randomly removing halo stars. The predominance of the trailing arm reflects the sky coverage of our sample,
as the leading arm lies largely outside the southern footprint of SkyMapper.

To quantify the residual contamination level and compare the removal 
completeness between the N-side and F-side on an equal footing, we 
count the stars in our final halo sample that satisfy 
$|\tilde{B}_\odot| \leq 10\degr$ but do not meet the distance criteria of 
\citet{2019MNRAS.486..378L}. These stars lie close to the Sgr 
stream orbital plane in projection, but their heliocentric distances 
are inconsistent with the stream distance at their sky position. As 
noted in Section~\ref{sec:data}, the distance criterion is precisely 
what prevents such stars from being over-cleaned, and most of them are 
therefore genuine halo stars rather than Sgr members. They consequently 
represent a conservative upper limit on residual Sgr contamination. We 
define the residual fraction as:
\begin{equation}
f_{\rm residual} = \frac{N_{\rm residual}}
{N_{\rm removed} + N_{\rm residual}}
\label{eq:fresidual}
\end{equation}
where $N_{\rm removed}$ is the number of Sgr members identified and 
removed using the distance criteria of \citet{2019MNRAS.486..378L} 
(Section~\ref{sec:data}), and $N_{\rm residual}$ is the number of 
residual candidates defined above. This definition gives the fraction of stars within 
$|\tilde{B}_\odot| \leq 10\degr$ that were not flagged as Sgr members, 
and can be compared directly between the two sides without being 
affected by their different sample sizes.

\begin{figure}[htbp]
   \centering
   \includegraphics[width=0.4\textwidth]{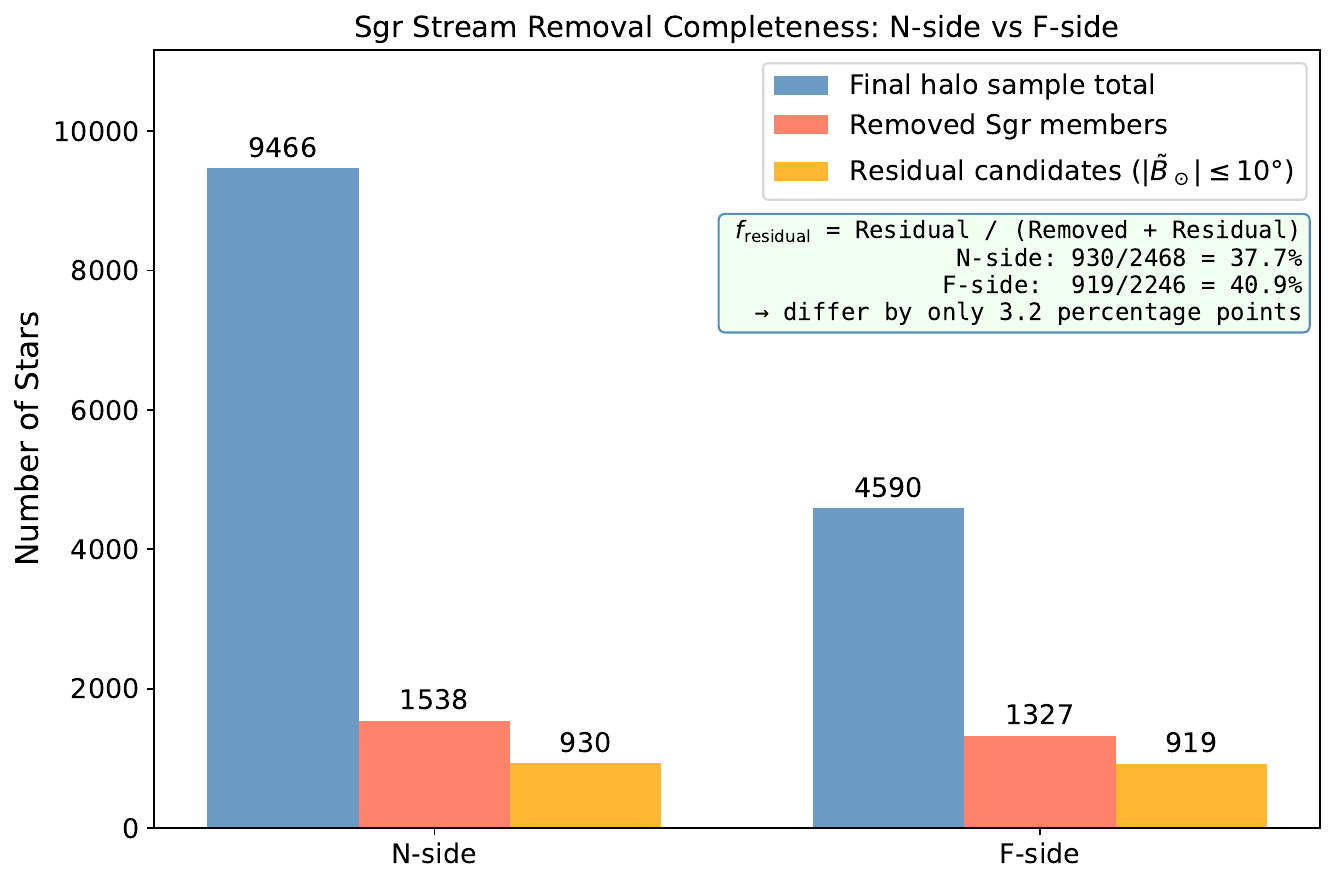}
   \caption{Quantitative assessment of the Sgr stream removal 
   completeness for the N-side and F-side subsamples. 
   Blue bars show the total number of stars in the final 
   halo sample on each side. Red bars show the number of 
   Sgr stream members removed by the criteria of 
   \citet{2019MNRAS.486..378L} and \citet{2017ApJ...850...96H}. 
   Orange bars show the number of residual Sgr candidates, 
   defined as stars in the final halo sample with 
   $|\tilde{B}_\odot| \leq 10\degr$ that do not satisfy the 
   \citet{2019MNRAS.486..378L} distance criteria; most of these 
   are genuine halo stars that happen to project close to the 
   stream plane. 
   The residual fraction $f_{\rm residual}$ (Equation~\ref{eq:fresidual}) 
   is 37.7\% on the N-side and 40.9\% on the F-side, 
   differing by only 3.2 percentage points, demonstrating 
   comparable removal completeness on both sides.}
   \label{fig:sgr_completeness}
\end{figure}

On the N-side, $N_{\rm removed} = 1{,}538$ and 
$N_{\rm residual} = 930$, giving $f_{\rm residual} = 930/2{,}468 
= 37.7\%$. On the F-side, $N_{\rm removed} = 1{,}327$ and 
$N_{\rm residual} = 919$, giving $f_{\rm residual} = 919/2{,}246 
= 40.9\%$. The two fractions differ by only 3.2 percentage points
(Figure~\ref{fig:sgr_completeness}), indicating that the removal procedure
achieves comparable completeness on both sides of the Galactic Center, with no
systematic bias toward either side. For completeness, we also express the
residual contamination relative to the final sample on each side. This gives
$930/9{,}466 = 9.8\%$ on the N-side and $919/4{,}590 = 20.0\%$ on the F-side.
This proxy is larger on the F-side largely because its final sample is smaller,
while the absolute numbers of residual candidates are nearly identical on the
two sides (930 versus 919). Any residual Sgr contamination therefore affects the
N-side and F-side equally and cannot introduce a spurious asymmetry or symmetry
between them.

\begin{figure*}[htbp]
   \sidecaption
   \includegraphics[width=0.62\textwidth]{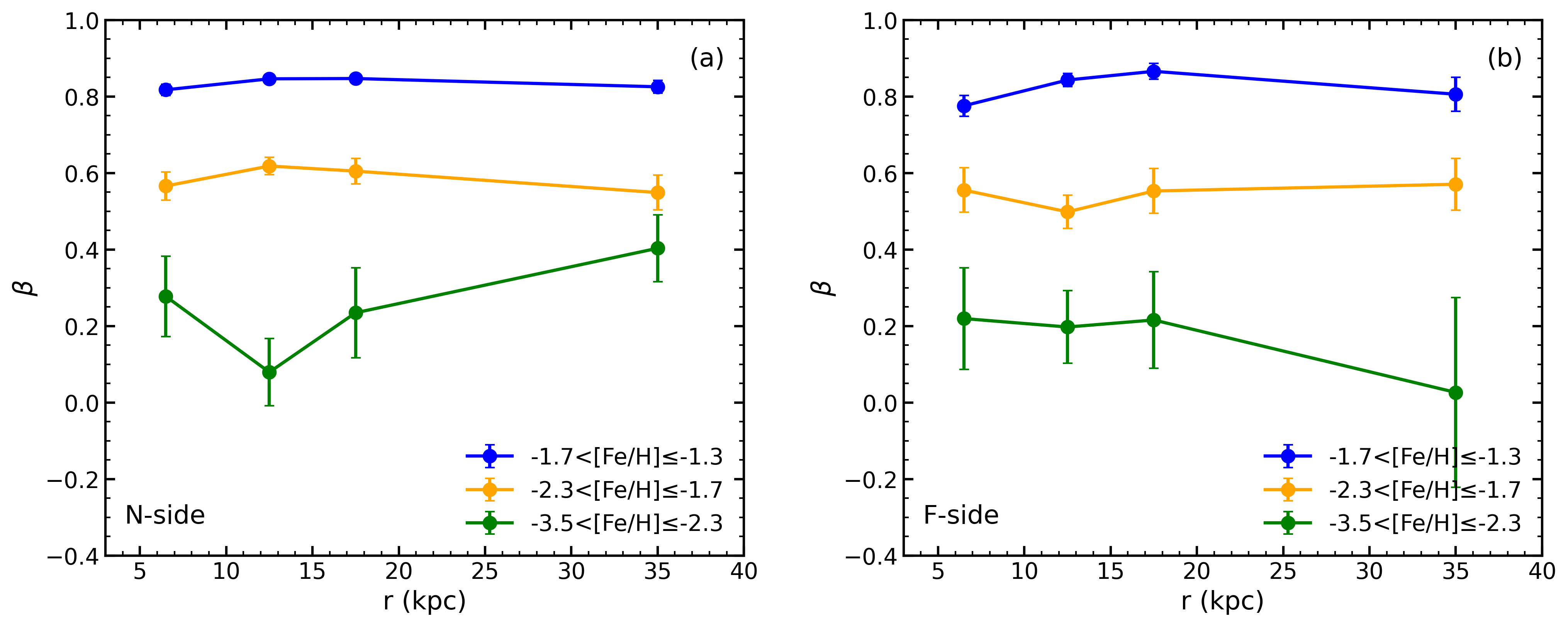}
   \caption{Velocity anisotropy parameter $\beta$ as a function 
   of Galactocentric distance $r$ for the conservative validation 
   sample, constructed by additionally removing all stars with 
   $|\tilde{B}_\odot| \leq 10\degr$ from the final halo sample. 
   Panel (a): N-side. Panel (b): F-side. 
   Blue, orange, and green curves correspond to the metal-rich 
   ($-1.7 < \mathrm{[Fe/H]} \leq -1.3$), metal-intermediate 
   ($-2.3 < \mathrm{[Fe/H]} \leq -1.7$), and metal-poor 
   ($-3.5 < \mathrm{[Fe/H]} \leq -2.3$) populations, respectively. 
   Error bars represent the $1\sigma$ uncertainty derived from 
   1000 bootstrap resamples. 
   The metal-rich population maintains $\beta \sim 0.8$ -- $0.9$ 
   on both sides, fully consistent with the main results 
   (Figure~\ref{fig:N_F_side_beta}), confirming that the GES 
   kinematic signatures are robust against residual Sgr stream 
   contamination.}
   \label{fig:beta_robust}
\end{figure*}

To directly verify that residual Sgr contamination does not affect 
our conclusions, we construct a conservative validation sample by 
additionally removing all stars with $|\tilde{B}_\odot| \leq 10\degr$ 
from our final halo sample, thereby eliminating every star that 
could conceivably be a residual Sgr member regardless of its 
distance. We emphasize that this step is performed solely as a 
robustness test and does not form part of our main analysis. Using 
this conservative sample, the velocity anisotropy parameter $\beta$ 
for the metal-rich population ($-1.7 < {\mathrm [Fe/H]} \leq -1.3$) 
remains at $\beta \sim 0.8$ -- $0.9$ on both the N-side and F-side 
(Figure~\ref{fig:beta_robust}), fully consistent with the main 
results. We therefore conclude that the GES kinematic signatures 
and their spatial symmetry reported in this work are robust against 
residual Sgr stream contamination.

\section{Conclusions}

In this study, we present a systematic investigation into the properties and
spatial symmetry of the remnants of the GES merger within the Galactic inner
halo, utilizing a sample of 14,056 red giant stars derived from Gaia DR3 and
SMSS DR4. By comparing the chemo-kinematic characteristics between the N-side
and F-side hemispheres, and further across four spatial quadrants, we are able
to test for the first time, with red giants as tracers, whether the imprint of
the GES extends symmetrically to the side of the Galaxy opposite the Sun.

Our central result is that the F-side bears a clear signature of the GES and is
symmetric to the well-studied N-side. Chemically, the F-side hosts the same
metal-rich component peaking at $\mathrm{[Fe/H]} \sim -1.4$ that characterizes
the GES debris, and dynamically it reproduces the canonical
``sausage''-shaped distribution in the $v_r$--$v_\phi$ plane. The F-side
exhibits an elevated mean prograde rotation relative to the N-side. However,
through a series of robustness tests in which the vertical-height cut is
progressively tightened, we show that this signal does not collapse toward zero
and therefore cannot be ascribed to residual disk contamination, but instead
reflects a genuine kinematic signal. Because $\beta$ characterizes the shape of
the velocity ellipsoid rather than its net rotation, this residual rotation does
not affect our symmetry diagnostics.

The evidence for the global nature of the GES comes from the velocity
anisotropy. The metal-rich population maintains $\beta \sim 0.8$ to $0.9$ within
the inner halo ($r < 28$~kpc) on both sides of the GC, while $\beta$ decreases
monotonically toward isotropy as the metallicity decreases. This chemo-kinematic
coupling is reproduced consistently on the N-side and the F-side, and our
four-quadrant analysis further shows that the high radial anisotropy of the
metal-rich component is azimuthally invariant, with $\beta$ spanning $\sim 0.6$
to $0.9$ across all quadrants throughout the inner halo. The action-space
analysis provides a complementary, quantitative confirmation. The fractions of
stars falling within the GES dynamical fingerprint region are
$11.56\%\pm0.16\%$ on the N-side and $10.86\%\pm0.23\%$ on the F-side, which a
two-proportion $z$-test ($z = 1.22$, $p = 0.22$) finds to be statistically
indistinguishable. Taken together, these results establish that GES debris does
not merely populate the Solar Neighborhood but fills the entire inner-halo
volume in a spatially symmetric and azimuthally uniform manner, with the
metal-rich, radially anisotropic GES component and a more isotropic, metal-poor
background coexisting throughout.

We have subjected this conclusion to extensive scrutiny. The N/F-side division is shown to be insensitive to the adopted $R_0$: varying it across the full range of literature determinations reassigns
at most $\sim6\%$ of stars, all near the boundary, and alters the
metal-rich $\beta$ by at most $\sim0.04$. A direct error-injection experiment
confirms that the comparable, and only marginally larger, measurement
uncertainties on the F-side can neither create nor distort the radial
anisotropy: inflating the N-side radial-velocity uncertainties to F-side
levels shifts $\beta$ by at most $3 \times 10^{-4}$, far below its
statistical uncertainty. Measurement errors can in any case only bias $\beta$
toward isotropy rather than enhance it. Finally, a quantitative assessment of the Sgr stream removal shows comparable completeness on both sides and demonstrates,
using a conservative validation sample, that residual stream contamination
cannot account for the observed symmetry. The N-side and F-side symmetry is
therefore a genuine physical property of the GES remnant rather than an artifact
of sample selection, measurement uncertainties, or substructure contamination.

In conclusion, by characterizing the distal Galactic halo and performing
multi-quadrant comparisons, this study removes potential biases arising from
local observational selection effects and corroborates the galaxy-wide dominance
and spatial symmetry of the GES remnants. These results provide observational
constraints for reconstructing the early assembly history of the MW and for
understanding the global dynamical evolution of its stellar halo. Looking
forward, with the release of subsequent Gaia data and deeper large-scale
spectroscopic surveys, we expect to further refine the timing of the GES
accretion and to characterize its perturbing effects on the Galactic disk
through more detailed chemical abundance analyses.

\begin{acknowledgements}
     We thank the anonymous referee for their careful reading and constructive comments, which have improved the clarity and rigor of this paper. This work is supported by the National Natural Science Foundation of China (NSFC, grant No. U1731108). Y.H. acknowledges the support from the National Key R\&D Programme of China (grant No. 2025YFF0510603), the National Science Foundation of China (NSFC grant No. 12422303), and the Fundamental Research Funds for the Central Universities (grant Nos. 118900M122, E5EQ3301X2, and E4EQ3301X2). T.C.B. acknowledges partial support from grants PHY 14-30152; Physics Frontier Center/JINA Center for the Evolution of the Elements (JINA-CEE), and OISE-1927130; The International Research Network for Nuclear Astrophysics (IReNA), awarded by the US National Science Foundation, and DE-SC0023128CeNAM; the Center for Nuclear Astrophysics Across Messengers (CeNAM), awarded by the U.S. Department of Energy, Office of Science, Office of Nuclear Physics.
     
     The national facility capability for SkyMapper has been funded through ARC LIEF grant LE130100104 from the Australian Research Council, awarded to the University of Sydney, the Australian National University, Swinburne University of Technology, the University of Queensland, the University of Western Australia, the University of Melbourne, Curtin University of Technology, Monash University and the Australian Astronomical Observatory. SkyMapper is owned and operated by The Australian National University's Research School of Astronomy and Astrophysics. The survey data were processed and provided by the SkyMapper Team at ANU. The SkyMapper node of the All-Sky Virtual Observatory (ASVO) is hosted at the National Computational Infrastructure (NCI). Development and support of the SkyMapper node of the ASVO has been funded in part by Astronomy Australia Limited (AAL) and the Australian Government through the Commonwealth's Education Investment Fund (EIF) and National Collaborative Research Infrastructure Strategy (NCRIS), particularly the National eResearch Collaboration Tools and Resources (NeCTAR) and the Australian National Data Service Projects (ANDS).
     
      This work has used data from the European Space Agency (ESA) Gaia mission (https://www.cosmos.esa.int/gaia), processed by the Gaia Data Processing and Analysis Consortium (DPAC, https://www.cosmos.esa.int/web/gaia/dpac/consortium).
\end{acknowledgements}

\bibliographystyle{aa} 

\bibliography{references}

\appendix
\nolinenumbers
\section{Sensitivity of the N/F-side separation to the adopted Galactocentric distance \texorpdfstring{$R_0$}{R0}}
\label{app:r0}

The division of our sample into the N- and F-side hemispheres is
defined by the $X = 0$ plane, computed with $R_0 = 8.178$~kpc
\citep{2019A&A...625L..10G}. The measured value of $R_0$ still carries a
considerable uncertainty \citep{2016ARA&A..54..529B}, and since the
dividing plane depends on $R_0$, we tested whether the adopted value
affects the hemisphere assignment and the resulting kinematics. We
repeated the full analysis over the entire range of $R_0$ values
reported in the literature, and re-divided the sample at $X = 0$ using
$X = -R_0 + d\cos l \cos b$ for each value, so that both the hemisphere
assignment and the kinematic quantities remain self-consistent with the
adopted $R_0$. In what follows we highlight the two extremes of this
range, $R_0 = 7.1$ and $8.92$~kpc.

\begin{figure*}[htbp]
\sidecaption
\includegraphics[width=0.62\textwidth]{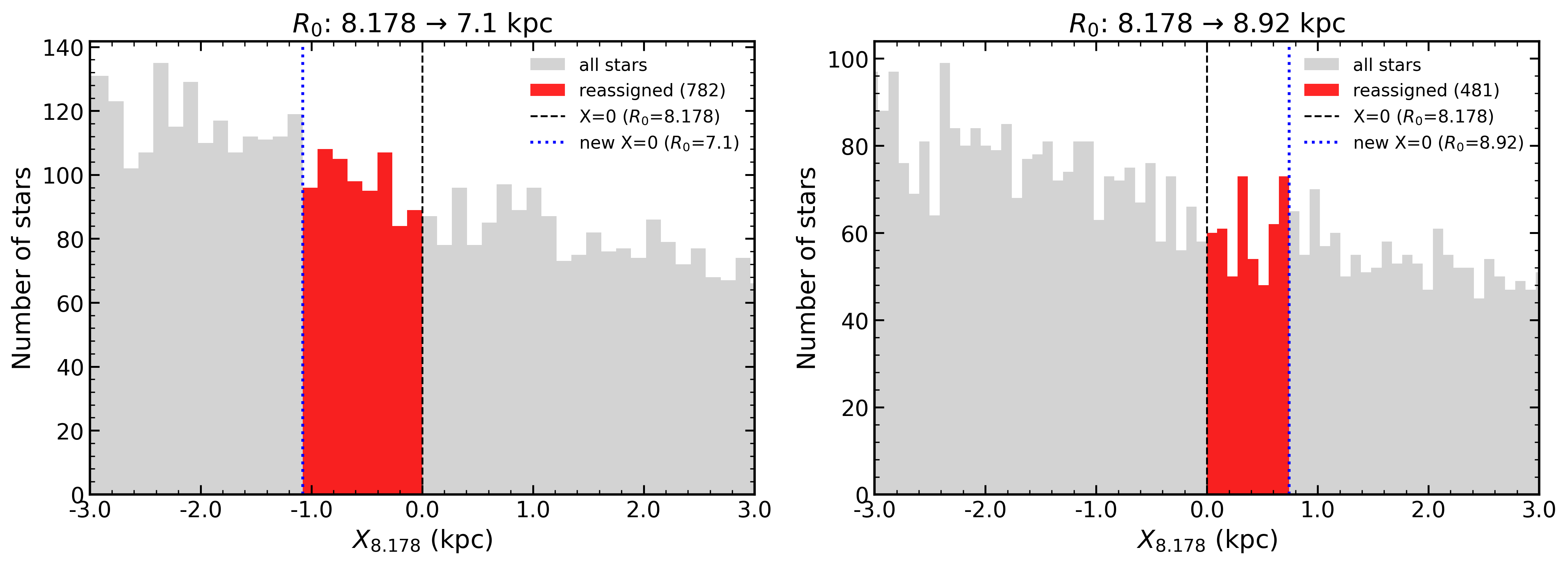}
\caption{Stars reassigned between the N- and F-side for the two extreme
$R_0$ values, shown in a zoom around the $X = 0$ plane. Left:
$R_0 = 8.178 \rightarrow 7.1$~kpc; right:
$R_0 = 8.178 \rightarrow 8.92$~kpc. Gray: all stars; red: reassigned
stars. The dashed line marks $X = 0$ for $R_0 = 8.178$~kpc, and the
dotted line marks the shifted $X = 0$ plane for the alternative $R_0$.
The reassigned stars are confined to the layer between the two planes
($|X| < 1.1$~kpc).}
\label{fig:r0_cross}
\end{figure*}

\begin{figure*}[htbp]
\centering
\includegraphics[width=0.8\textwidth]{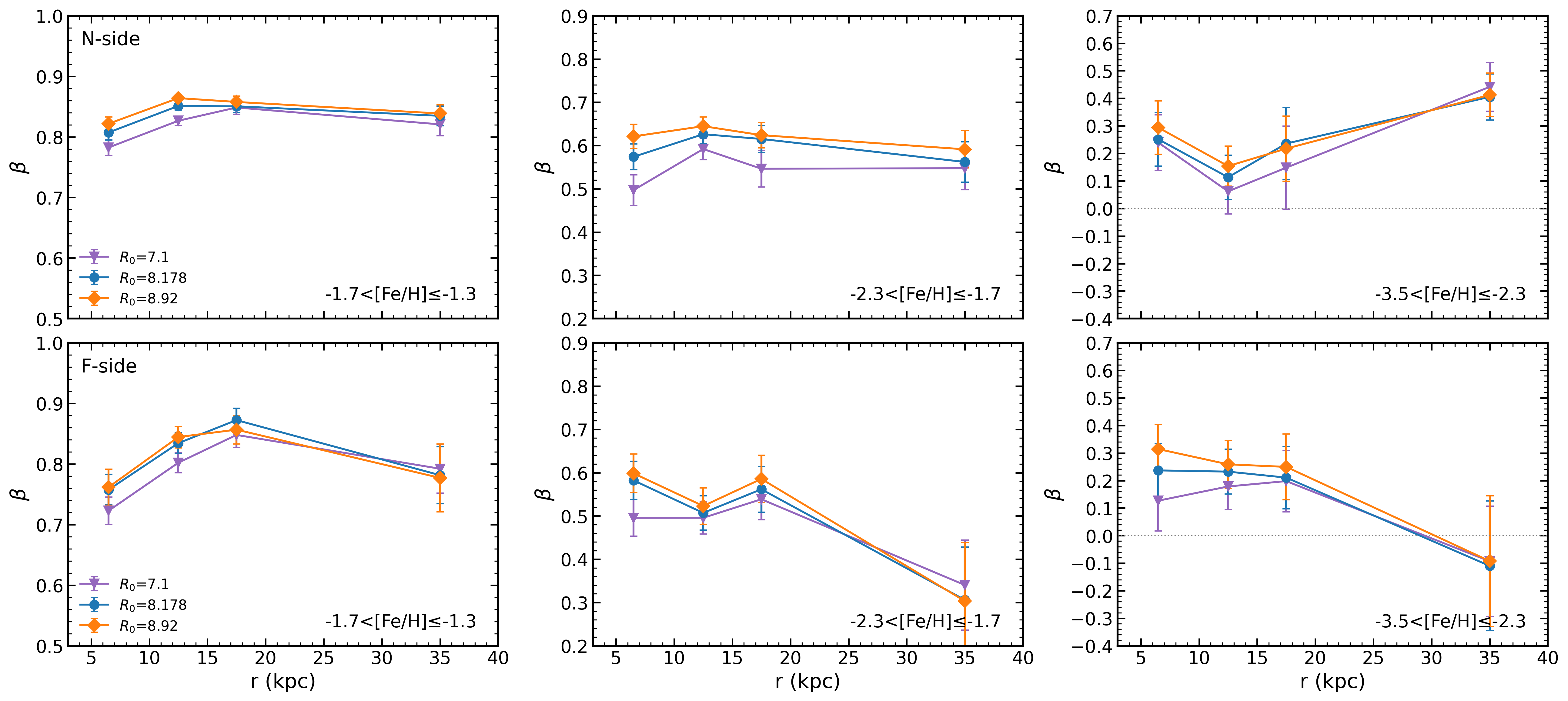}
\caption{Velocity-anisotropy parameter $\beta$ as a function of
Galactocentric radius $r$ for $R_0 = 7.1$ (purple), 8.178 (blue), and
8.92~kpc (orange), i.e. the two literature extremes and our adopted
value. Top row: N-side; bottom row: F-side. Columns from left to
right: metal-rich ($-1.7 < \mathrm{[Fe/H]} \le -1.3$),
metal-intermediate ($-2.3 < \mathrm{[Fe/H]} \le -1.7$), and metal-poor
($-3.5 < \mathrm{[Fe/H]} \le -2.3$) populations. Error bars are
$1\sigma$ from 1000 bootstrap resamples. For the metal-rich (GES)
population, $\beta$ differs by at most $\sim0.04$ among the three $R_0$
values, demonstrating that the kinematic structure on both sides is
insensitive to the adopted $R_0$.}
\label{fig:r0_beta}
\end{figure*}

Even for these extreme values, which differ from our adopted $R_0$ by
up to $\sim1.1$~kpc, only a modest fraction of stars are reassigned
between the two hemispheres: 782 stars (5.6\%) and 481 stars (3.4\%),
respectively, all confined to the layer between the old and new
$X = 0$ planes ($|X| < 1.1$~kpc; Fig.~\ref{fig:r0_cross}). As a result,
the kinematics are essentially unchanged (Fig.~\ref{fig:r0_beta}):
across the full range the velocity-anisotropy parameter $\beta$ of the
metal-rich (GES) population on both sides differs by at most
$\sim0.04$, comparable to or within the bootstrap uncertainties, and
the radially dominated signature ($\beta \approx 0.8$) in the inner
halo, including the innermost radial bin, is fully preserved. We
therefore conclude that our results are robust against the adopted
value of $R_0$ over the entire range of literature values.

\end{CJK}  

\end{document}